\documentclass[10pt,aps,nofootinbib,superscriptaddress,twocolumn,preprintnumbers,balancelastpage]{revtex4-1}
\usepackage{amsmath,amssymb,amsthm,amsopn}
\usepackage{epsfig}
\usepackage{psfrag}
\usepackage[colorlinks=true,allcolors=blue!65]{hyperref}
\usepackage{multirow}
\usepackage{empheq}
\usepackage{slashed}
\usepackage{graphicx}
\usepackage{url}
\usepackage{subfigure}
\usepackage{textcomp}
\usepackage{bm}
\usepackage{dcolumn}
\usepackage{color,xcolor}
\usepackage{ulem}
\usepackage{cancel}
\usepackage{braket}
\usepackage[title]{appendix}
\usepackage{lipsum}
\usepackage{mathptmx}

\DeclareMathAlphabet{\mathcal}{OMS}{cmsy}{m}{n}

\def\nn{\nonumber}

\def\comment#1{}

\def\beq{\begin{equation}}
	\def\eeq{\end{equation}}
\def\bea{\begin{eqnarray}}
	\def\eea{\end{eqnarray}}

\begin{document}
	\renewcommand{\topfraction}{1.0}
	\renewcommand{\bottomfraction}{1.0}
	\renewcommand{\textfraction}{0.0}
	
	\newcommand\sect[1]{\emph{#1}---}

	\title{Anomalous dimension and quasinormal modes of flavor branes}
	\author{Mahdi Atashi}
	\email{m.atashi@shahroodut.ac.ir}
	\affiliation{Faculty of Physics, Shahrood University of Technology, P.O.Box 3619995161 Shahrood, Iran}
	
	\author{Kazem  Bitaghsir  Fadafan}
	\email{bitaghsir@shahroodut.ac.ir}
	\affiliation{Faculty of Physics, Shahrood University of Technology, P.O.Box 3619995161 Shahrood, Iran}

	\begin{abstract}
	We study scalar quasinormal modes in a D3/D7 system holographically dual to a quantum field theory with chiral symmetry breaking at finite temperature. From the bottom-up approach, we consider a nontrivial dilaton profile which is responsible for the anomalous dimension of the quark condensate. It depends on a new parameter$ q$ in the model. By varying this parameter, we study the behavior of the massive and massless scalar quasinormal modes. The numerical method that we use is the spectral method, and we find that there is no pure imaginary mode for the massless case but it appears by increasing the parameter $q$.  It is known that this mode becomes tachyonic for massive cases. Then we turn on a pseudoscalar field and using a simple ansatz study its effect on the quasinormal modes of the scalar field. By varying the parameter of the nontrivial dilaton profile in the model, we qualitatively study quasinormal modes in walking theories.
			 
	\end{abstract}
	
	\maketitle

	\section{Introduction}

	The AdS/CFT correspondence provides a useful tool to study quark-gluon plasma (QGP) produced at LHC and RHIC \cite{Casalderrey-Solana:2011dxg}. It relates the strongly coupled non-Abelian gauge theory to a dual classical gravity. A well known example is the duality between $ \mathcal{N}=4 $ SYM and type IIB superstring theory on the AdS$_5 \times $S$^5$ \cite{Aharony:1999ti}. To consider finite temperature gauge theory, one should add a black hole to the bulk geometry. Although a precise holographic dual to QCD is not known, several important transports have been found, like shear viscosity over entropy density, the energy loss of heavy quarks or even non-equilibrium phenomena and fast thermalization which has been discussed in \cite{Kovtun:2004de}. For studying quantum or coupling corrections from the holographic point of view see \cite{Fadafan:2008gb, BitaghsirFadafan:2010rmb,Grozdanov:2016vgg, Atashi:2016fai}.
	
	By embedding D7 branes in AdS$_5 \times $S$^5$, new degrees of freedom in the fundamental representation is added to the theory \cite{Karch:2002sh}. This can be done in the quench limit where the number of D7 branes $N_f$ is less than the number of D3 branes $N_c$ in the background, then we are ignoring the backreaction of D7 branes. This is called the D3/D7 setup and the study of mesons in this setup has been reviewed in \cite{Erdmenger:2007cm}. There are two main types of embeddings for D7 branes in the background of D3 branes, Minkowski and black hole embeddings \cite{Mateos:2006nu}. The dual theory is given by $ \mathcal{N}=4 $ SYM $SU(N_c)$ theory at finite temperature, $T$ with $ \mathcal{N}=2 $ hypermultiplets with mass $m$ in the fundamental representation. In the dual field theory, the dimensionful parameters are $T$ or $m$ and one can set either of them to be one. In this way, lowering mass means raising $T$. From the bulk point of view, high $m$ corresponds to Minkowski embeddings of D7 branes while low $m$ refers to black hole embeddings. Thus Minkowski and black hole embeddings should be considered at low temperatures and high temperatures, respectively. In this paper we study the fluctuations of the black hole embeddings.
	
	Mesons are holographically modeled by open strings ending on the D7 probe branes \cite{Erdmenger:2007cm}. Low spin mesons are found by studying fluctuations of the D7 brane around the equilibrium. Then to study them one has to consider D7 brane fluctuations for both Minkowski and black hole embeddings. In the former one finds a meson mass spectrum because the fluctuation wave travels along the D7 brane and will reflect at the end point of it. As a result, one gets a discrete spectrum of meson masses. However, in the latter case, the D7 brane falls into the black hole and one has to impose infalling boundary conditions for fluctuation wave at the horizon. As a result, one finds complex frequencies for eigenmodes which are called quasinormal modes (QNMs) \cite{Hoyos-Badajoz:2006dzi}. From the AdS/CFT correspondence, they are mapped to the poles of the retarded correlation functions of the dual theory at finite temperature \cite{ Kovtun:2005ev}. The imaginary part of a QNM is related to the relaxation time of a small fluctuation of the medium around its equilibrium. Here on the D7 brane, QNMs represent the melting of mesons in the plasma at high temperatures. For the Minkowski embedding, mesons are stable and will not be melted by such perturbations. However, by increasing the temperature the system goes through a phase transition, represented in the bulk by a transition between Minkowski and black hole embeddings. The decay of mesons and QNMs at high temperatures have been studied in \cite{Hoyos-Badajoz:2006dzi}.
	
	Recently, the D3/D7 setup has been adjusted phenomenologically by considering a running anomalous dimension $\gamma$ for the condensate of quarks, using an effective description of a dilaton profile \cite{BitaghsirFadafan:2019ofb}. Interestingly, it supports hybrid compact stars with quark cores, see a recent review of holographic neutron stars \cite{Jarvinen:2021jbd}. The running of $\gamma$ allows us to describe the chiral symmetry restoration transition away from the deconfined phase of massive quarks. This transition occurs for the scalar dual to the chiral condensate when it violates the Brietenlohner Freedman (BF) bound \cite{BF}. The running of $\gamma$ could be controlled in the model by the derivative of it at the point that BF bound violates.
	
	Such D3/D7 setups are dulal to QCD-like gauge theories so that we consider $SU(N_c)$ Yang-Mills theory coupled to $N_f$ Dirac fermions transforming in the fundamental representation. Then the theory can be studied in the plane of $(N_c,N_f)$. The effective dilaton profile provides a free parameter to explore this plane. These QCD-like theories correspond to the UV free gauge theories with a matter in the fundamental representation. Interestingly, they show a conformal window in the parameter space of $N_c$ and $N_f$ which flows to a CFT in the IR. In the quenched limit, one treats $x=\frac{N_f}{N_c}$ as a continuous parameter and finds the window at $x_c<x<11/2$. At $x=11/2$, the beta function vanishes at one loop which can be studied perturbatively. However, studying the lower edge of the window is hard so different methods have been considered \cite{DiPietro:2020jne}. At critical $x_c$, the IR conformal theory is replaced by one with chiral symmetry breaking in the IR with a mass gap. When the anomalous dimension of quark condensate hits order one, the chiral symmetry breaking is triggered in these theories. It is expected that below $x_c$ the theory shows walking behavior where over the long range of energy the coupling constant varies slowly. One should notice that lattice methods do not show a perfect agreement on the value of $x_c$ \cite{DeGrand:2015zxa, Hasenfratz:2019dpr}. As a useful tool, we use the AdS/CFT correspondence to study such QCD-like gauge theories. Using this new approach, the Veneziano limit and the walking regime have been studied in \cite{Kutasov:2012uq, Jarvinen:2011qe, Alvares:2012kr, Jarvinen:2015qaa}, and \cite{Alho:2013dka, BitaghsirFadafan:2018efw}, respectively. In this study, we are going to use the adapted D3D7 with the dilaton profile to investigate the related physics of QCD-like gauge theories.

In this paper we compute QNMs of this adapted D3/D7 setup and investigate how they change in the presence of running anomalous dimension. By changing the free parameter in the effective dilaton profile, we explore indirectly the parameter space in the plane of $(N_C,N_f)$. As it was expressed, this is a bottom up phenomenological model, and there is no precise relation between $q$ and  $x=\frac{N_f}{N_c}$ \cite{Alvares:2012kr}. However, we will show that by varying $q$, we are also changing the anomalous dimension of the quark condensate. By computing the QNMs of the scalar field in the adapted bottom up D3/D7 setup, we are studying the meson melting. As it was expressed, there are two different D7  embeddings and a first order phase transition occurs between them \cite{Mateos:2006nu}. It is found that a tachyonic mode shows up in this scalar sector of the QNMs which is close to the first order phase transition \cite{Kaminski:2009ce}. By varying $q$ parameter, we investigate how this mode is changing regarding the mass of the quarks in the system. We will discuss how the black hole D7 probe brane embedding is changing in this case.
	
First, we consider flat D7 brane embedding corresponding to massless quarks. Next we do the calculations for massive quarks and consider both vanishing and finite momentum for the fluctuations. We will present the results for the first ten QNMs. Although we do not compute the free energy of the system and the phase transition but as \cite{Kaminski:2009ce} we show that how an instabilty occurs in the system. We find an excellent numerical agreement between maximum quark mass (Figure \ref{Fig1} ) and purely imaginary QNMs (Figure \ref{Fig4} ). In this way, we explicitly show where the pure imaginary modes cross the real axis and system becomes unstable.
	
 It is important to mention that unlike \cite{Hoyos-Badajoz:2006dzi} and \cite{Kaminski:2009ce}, in this paper, we employ the spectral method to solve the fluctuation equations. The exponential convergence of the spectral method gives it greater rapidity and accuracy than other methods such as shooting, relaxation, and the Leaver method. We could compute QNMs in less than one minute for massless and around 40 minutes for massive cases by a PC with $ i7-6700 $ CPU$@$3.4 GHz and $ 16 $GB RAM, those are much less than other methods. One finds details of the spectral method in \cite{Jansen:2017oag}.
	\section{Holographic bottom-up D3/D7 model with chiral symmetry breaking}
	  In this section we use the D3/D7 setup and construct a holographic bottom-up D3/D7 model with an explicit chiral symmetry-breaking mechanism, one finds more details in \cite{Alvares:2012kr}.\footnote{Our convention for coordinate $\rho$ is not the same as \cite{Alvares:2012kr}.} A precise top-down D3/D7 example has been studied in \cite{Filev:2007gb} where the chiral symmetry breaking mechanism is triggered by a magnetic field.

	We consider the D7 brane black hole embedding solution corresponding to the finite temperature description of the dual field theory. The $AdS _5 \times \text{S}^5 $ background metric is
		\begin{eqnarray}\label{bulkmetric}
			ds^2& =& \frac{\rho ^2}{2 L^2}\left(-\frac{f^2}{\tilde{f}} ~dt^2 +\tilde{f} ~d\vec{x}^2\right)\\\nn &&+ \frac{L^2}{\rho^2 } \left(dr^2+r^2 ds_{S^3}^2+dR^2+R^2d\psi^2 \right) \ ,
		\end{eqnarray}
		where $ds_{S^3}^2$ is the metric of a unit radius $S^3$. Also, $r$ runs over the D7 brane coordinates and the transverse coordinates to both D3 and D7 branes denote as $R(r)$ and $\psi(r)$. The radial coordinate is given by $ \rho^2=r^2+R(r)^2 $ and from the AdS/CFT dictionary $L^4=4 \pi g_s N \alpha^{'2}$ where $L$ is the AdS radius. The metric functions are given by
		\begin{eqnarray}
			f(\rho) = 1-\frac{\rho_H^4}{\rho^4}, \qquad 	\tilde{f}(\rho) = 1+\frac{\rho_H^4}{\rho^4}.
		\end{eqnarray}
		Here $\rho_H$ is the position of the black hole horizon which is
		related to the temperature as $ T=\frac{\rho_H}{\pi}L^2 $, and the boundary is located at $ \rho \to \infty $. We set $ L=1 $ and $ T=1/\pi $ from now on.

		As it was mentioned, the bottom-up model is based on the D7 brane embedding with Dirac-Born-Infeld (DBI) action. The scalar field $R(r)$ is dual to the chiral condensate in the boundary theory. In the quenched limit ($N_f \ll N_c$), ${\cal N}$=2 quark superfields are included in the ${\cal N}$=4 gauge theory through these probe D7 branes. The D3-D7 connected open strings describe the quarks while D7-D7 strings holographically describe mesonic operators and their sources. 
		
		 We add the chiral symmetry breaking mechanism in the model by considering a new dilaton profile function $h$
		\begin{equation}\label{profileh1}
				h(\rho)=1+\frac{C}{\rho^q},
		\end{equation}		
		 in the DBI action of the D7 branes
		\begin{eqnarray} \label{dbi1}
	 S_{DBI} = - T_{D7} ~ \int d^8\xi\,h(\rho) ~\sqrt{-\text{det P[G]}},
	\end{eqnarray}
	where P[G] is the pullback of the background metric $G_{ab}$ given in \eqref{bulkmetric} as 
	\begin{eqnarray} \label{dbi2}
		\text{P[G]}=\partial_{\mu}X^a ~G_{ab} ~\partial_{\nu}X^b.
	\end{eqnarray}
	Here $ \{a,b\} $ run over $ 10-$ dimensional background metric, $ \xi_{\mu} $ and $ X^a $ are the coordinate of D7 brane and the background coordinates, respectively. The constant $C$ is a dimensionful parameter and in units of temperature it is a free parameter in the model. We we set it to be one in this study. 
	
	Now we explain the important role of the function $h(\rho)$ in the model. It is an effective dilaton field. One should notice that the dilaton field in $\mathcal{N}=4$ SYM theory is constant and does not run by changing the scale of the energy but in this bottom-up model, it is a non-trivial function that triggers chiral symmetry breaking. There exists also some assumptions for choosing the appropriate function $h$. In the UV it returns a constant with $\gamma=0$ but runs to a fixed point value at low energy. The equation \eqref{profileh1} is a simple choice for $h$ that satisfies the conditions but finding the precise form of $h$ using a top-down model is not easy. Applying the precise model of the running coupling from QCD to the evolution of the anomalous dimension of the condensate presents complexities. Consequently, a tachyon field was examined in relation to the chiral condensate, and holographic beta functions were meticulously constructed as outlined in \cite{Jarvinen:2011qe}. The parameter $C$ is a free parameter and is a crucial scale in the context of the model and implies transition from weak to strong coupling. By setting $C=1$, one essentially normalizes the scale, which simplifies the analysis of the theory's behavior as it transitions between these coupling regimes. At finite temperature, we set it to the IR physics at the horizon.
	
		One obtains the D7 brane action in the background \eqref{bulkmetric} with $\psi(r)=0$, 
		\begin{eqnarray} \label{dbi1}
			S_{DBI} = - T_{D7} ~  \int \, \lambda(\rho)\,r^3\,d r ~\sqrt{1+R'(r)^2},
		\end{eqnarray}  
		where $\lambda(\rho)=h(\rho) f(\rho)\, \tilde{f}(\rho)$.
				
According to the duality, the scalar field $R(r)$ is related to the chiral symmetry breaking in the dual field theory. Its equation of motion in the background \eqref{bulkmetric} is given by
	
	\bea \label{eom1}
	 \partial_{r}\left(\frac{\lambda(\rho) r^3 R'(r)}{\sqrt{1+R'(r)^2}}\right)-2R(r) \frac{\partial \lambda(\rho)}{\partial \rho^2} r^3 \sqrt{1+R'(r)^2}=0.
	\eea
	The massless quark solution with zero quark condensate corresponds to an embedding solution $R(r)=0$ which is a solution of the \eqref{eom1}. But the flat embedding $R(r)=m$ is no longer a solution if in the second term $\frac{\partial \lambda(\rho)}{\partial \rho^2}$ be a non-trivial function of $R(r)$. Following \cite{Alvares:2012kr} by expanding the action,  when $\lambda$ is constant, one finds that the scalar field $R(r)$ maps to a field $\phi$ with $m^2=-3$. Here two fields are related with a new coordinate as $R=\hat{r} \phi$ and 
	\bea
	\hat{r}=\frac{1}{\sqrt{2}}\left(\int_r^\infty \frac{d r}{\lambda(\rho)\, r^3} \right)^{-\frac{1}{2}},
	\eea
	where canonical mass of new field $\phi$ is related to the $\lambda$ function as 
	\begin{equation}\label{deltam}
		m^2= -3 - \delta m^2,\,\,\,\,\, \delta m^2=\lambda(\rho)\,\lambda'(\rho)\, \frac{r^5}{\hat{r}^4}. 	
	\end{equation} 
	Where we take the derivative with respect to coordinate $r$. The function $\lambda{(\rho)}$ vanishes at horizon and by changing the radial coordinate, $m^2$ passes through $-4$ then the BF bound is violated and the D7 embedding moves away from the flat embedding $R=0$ and chiral symmetry breaking occurs in the dual theory. Using the mass operator dimension relation for the scalar field, $ m^2=\Delta (\Delta-4)$, one finds 
	\bea
	\gamma=1-\sqrt{1-\delta m^2},
	\eea
	where $\delta m^2$ is given in \eqref{deltam}. As it was explained, the BF bound is violated when $\gamma>1$. It is important to notice that we are considering the AdS black hole geometry as the background so that the dual theory is at a finite temperature. On the other hand, in \cite{Alvares:2012kr} the nontrivial dilaton profile, \eqref{profileh1}, has been proposed to interpret $q$ as a control parameter of anomalous dimension in the system. In our study, the IR physics is not the same as the zero temperature discussed in \cite{Alvares:2012kr} because of the presence of the black hole horizon in which the function $ \lambda(\rho)$ , which is related to the shift of mass squared, approaches zero at the horizon. However it does not change the interpretation of $q$. One should notice that the function $ h$ which is crucially a function of $r+R[r]$ is the key extra ingredient in our study.  In a complicated geometric it backreacted on the metric background but we did not consider such phenomena. Also we are using a bottom-up approach which allows us to consider the dilaton profile at finite temperature as well. Besides the main advantage of this particular profile is triggering chiral symmetry in the system. As we know, the relationship between chiral symmetry breaking and temperature is a significant aspect of QCD and has implications for understanding the behavior of hadronic matter under extreme conditions for example in heavy ion collisions. In the context of QCD, chiral symmetry breaking is a non-perturbative phenomenon that occurs at low energies. It is known that there is a critical temperature, denoted as $ T_c $, below which chiral symmetry is spontaneously broken, and a corresponding mass gap is dynamically generated. But at high temperatures the chiral symmetry is expected to be restored. This means that as the temperature increases and approaches $ T_c $ from below, the chiral symmetry breaking effects diminish and eventually disappear, leading to a phase where the symmetry is unbroken. As a result, our goal in this paper is studying the chiral symmetry breaking in the presence of temperature.  As mentioned choosing dilaton profile is coming from adding magnetic field in the DBI action. In this case breaking of the chiral symmetry has been studied in [18].  As a result, the interpretation of $q$ does not change and that would be interesting to study how the phase diagram of QCD at finite temperature and density changes for different values of this parameter. It is worth to mention that  parameter $q$ in \eqref{profileh1} is related to the physics of walking theories \cite{Alvares:2012kr}. For a simpler choice of $h=\frac{1}{r^q}$, it is found that at $q=0.536$, the anomalous dimension takes the BF bound while $q$ close to $2$ means theories which run quickly to large IR fixed point. For $q>2$, the theory has a divergent anomalous dimension at finite holographic distance $\rho$ dual to a finite energy scale. However, it is interesting that the dual gravity theory provides a smooth description below this scale. We explore the parameter space of the model by changing $q$ in \eqref{profileh1} corresponding to the dual walking theories. \\

	\section{Quasinormal modes}
In this section, we analyze the scalar QNMs for the case of D7 brane black hole embeddings. They are time-dependent perturbations that show damped oscillatory behavior and dissipate the energy into the black hole. From the dual theory description, one finds that at high temperatures they correspond to the melting of bound state mesons \cite{Hoyos-Badajoz:2006dzi}. It is much easier to do the numerical computations in the new coordinate $ u=\rho_H/\rho $.  In this coordinate, the horizon is located at $ u=1 $ and the boundary is at $u=0 $. Then one should replace $R(r)$ and $r$ with new coordinates $\theta(u)$ and $u$,
	\bea
\theta=\tan^{-1}\left(\frac{R}{r}\right),\,\,\,\,\,u= \frac{L^2}{\sqrt{r^2+R^2}} .
	\eea
	The background geometry \eqref{bulkmetric} in the new coordinate is given by
	\begin{eqnarray}\label{umetric}
		ds^2 &=& \frac{1}{u^2}\big(- g(u) ~dt^2 +\frac{1}{g(u)} ~du^2 + d\vec{x}^2 \big)\\\nn&&+d\theta^2+\sin^2\theta ~d\psi^2+\cos^2\theta ~ds^2_{S^3},
	\end{eqnarray}	
	where $ g(u)=1-u^4 $. The function $h$ in \eqref{profileh1} reads  
	\begin{eqnarray}
		h(u) = 1+u^q.
	\end{eqnarray}
	The D7 brane's world volume coordinates are $\left(t,x,y,z,u\right)$ plus the $S^3$ coordinates. They are embedded along $\text{AdS}_5\times S^3$ background. Now, the world volume scalar fields are $\theta(u)$ and $\psi(u)$ which determine the modulus $m$ and phase $\psi$ of the hypermultiplet quark mass, respectively.
	\begin{figure}[tbp]
	\includegraphics[width=8.5cm]{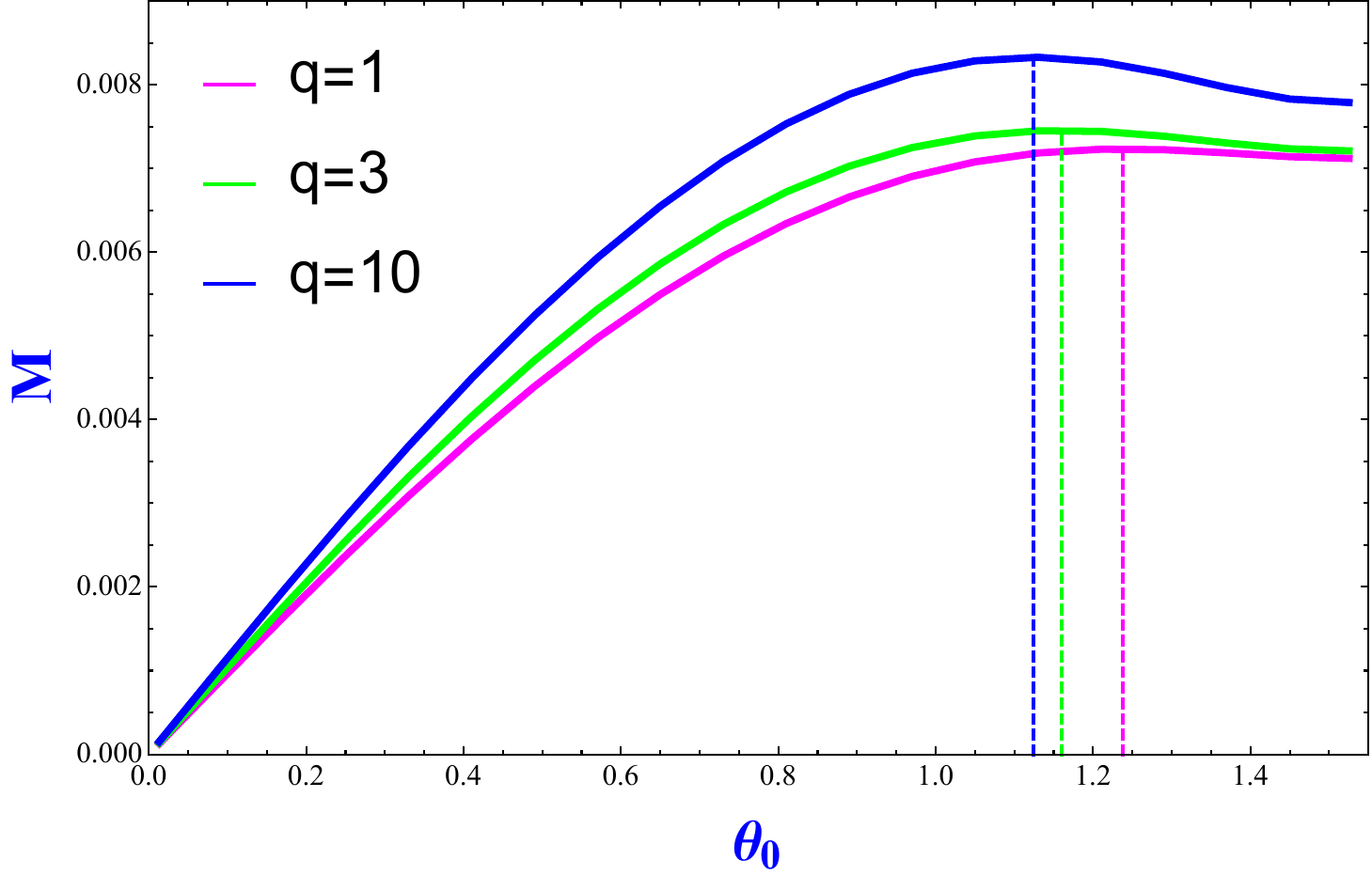}	
	\caption
	{ The mass parameter $M$ as a function of the embedding angle $\theta_0=\theta(u=1)$. We consider black hole embeddings so that the horizon is located at $ u=1 $ and the boundary is at $u=0 $. From top to bottom $q=10, 3$ and $q=1$, respectively. The vertical lines show the maximum quark mass in the system. For $q=10, 3, 1$, the embedding angles are $\theta_0=1.12,1.16,1.24$, respectively. These values are the same as the Figure \ref{Fig4} where the pure imaginary modes cross the real axis and system becomes unstable.}
	\label{Fig1}
\end{figure}
	In the new coordinates, the DBI action of the scalar field $\theta(u)$ is given by 	
	\begin{eqnarray}
		S_{D7}=\int du  ~h(u) \frac{\cos ^3 \theta(u)}{u^5} \sqrt{1+u^2 (1-u^4) ~\theta '(u)^2}
	\end{eqnarray}
	One obtains the equation of motion	
	\begin{eqnarray}\label{embedding}
		0 &=& \sin \theta (u) \nn \\
		&& \bigg (3 u^2 (u^4-1) (u^q+1) ~\theta '(u)^2  -3 (u^q+1)\bigg ) \nn \\
		& & +u\,\cos  \theta (u) \nn \\
		&&\bigg (\big((q+1) ~u^{q+4}-(q-3) ~u^q+u^4+3\big) ~\theta '(u) \nn \\
		& & -u^2 (u^4-1)\big((q-2) ~u^{q+4}-\nn \\ 
		&&(q-4) ~u^q-2 u^4+4\big) ~\theta '(u)^3 \nn \\
		& &  +u ~(u^4-1) (u^q+1) ~\theta ''(u)\bigg ).
	\end{eqnarray}
  \begin{figure*}[tbp]
	\centering
	{\includegraphics[width=14cm]{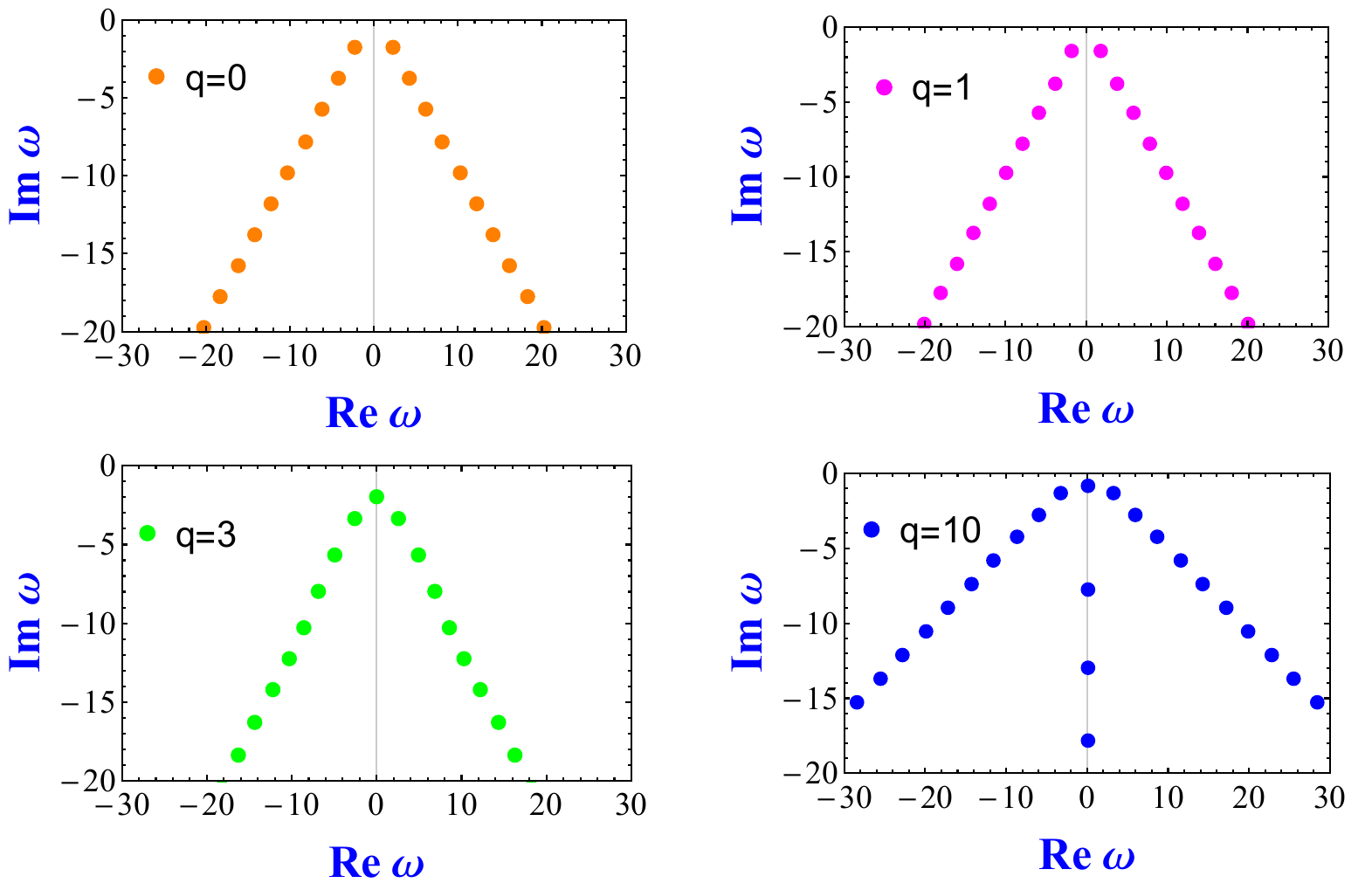}	}		
	\caption
	{Massless quasinormal frequencies at vanishing momentum for $ q=0 $ (top-left), $ q=1 $ (top-right), $ q=3 $ (bottom-left), and $ q=10 $ (bottom-right). By increasing $q$, the massless quasinormal modes start colliding in the imaginary axis and moving down in the complex frequency plane. For $q=10$, a tower of pure imaginary modes is clearly seen in this figure.
	}\label{Fig2}
\end{figure*} 
	\begin{figure}[tbp]
	\includegraphics[width=8.2cm]{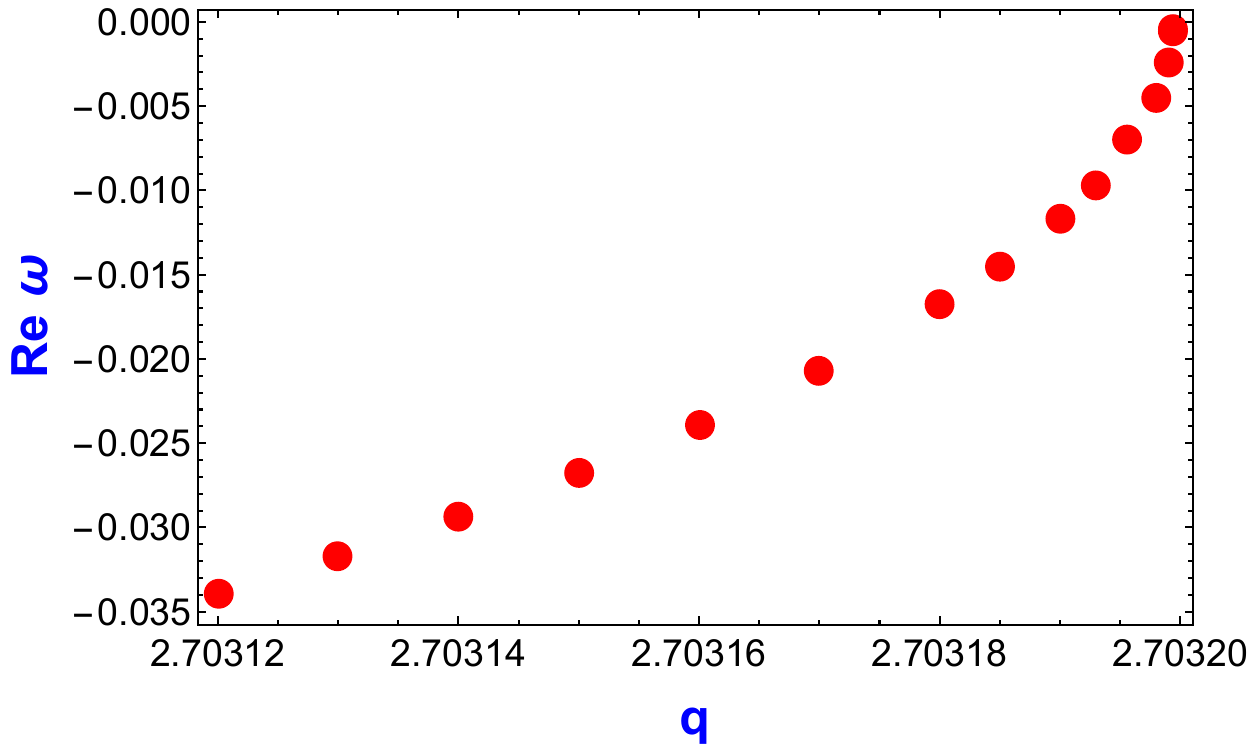}	
	\caption
	{ Real part of the firs quasinormal frequency versus the $q$ parameter. There is a critical $q_c \approx 2.70319$ in which the real part vanishes and the pure imaginary mode shows up as in Figure \ref{Fig2}.
	}\label{Fig2-2}
\end{figure}
At $q=0$, it is similar to \cite{Kaminski:2009ce} with a different definition of $\theta$ angle.
\begin{table}
	\begin{center}
		\begin{tabular}{|c||c|c|} \hline 
			n & $Re (\omega_n)$  &  $Im (\omega_n)$ \\ \hline \hline
			1 &2.198814566 & -1.759534627 i  \\ \hline
			2 &4.211897203 & -3.774888236 i  \\ \hline
			3 &6.215543149 & -5.777257013 i  \\ \hline
			4 &8.217167238 & -7.778080220 i  \\ \hline
			5 &10.21806124 & -9.778473884 i  \\ \hline
			6 &12.21861770 & -11.77869747 i  \\ \hline
			7 &14.21899318 & -13.77883893 i  \\ \hline
			8 &16.21926141 & -15.77893528 i  \\ \hline
			9 &18.21946137 & -17.77900453 i  \\ \hline
			10 &20.2196154 &  -19.7790564 i  \\ \hline
			11 &22.21974 & -21.77910 i  \\ \hline
			12 &24.220 & -23.779 i  \\ \hline
			13 &26.2 & -25.8 i  \\ \hline   
		\end{tabular}
	\end{center}
	\caption{ Thirteenth scalar massless QNMs computing by spectral method at $q=0$. We used 61 points in the grid which was judged by a comparison with a grid with 81 points. }
\end{table}

The D7 brane embeddings are found by solving this differential equation from the horizon $u=1$ out to the boundary $u=0$. Black hole embedding is found by imposing boundary condition $ \theta (1)=\theta_0 $ and $ \theta'(1)=3 \tan \theta_0/4 $ on (\ref{embedding}). Minkowski embedding is found by considering $ \theta(u_0)=\pi/2 $ and $ \theta'(u_0) \to \infty $. 

Figure \ref{Fig1} shows the effect of $q$ on the D7 brane black hole embeddings. In this figure, from top to bottom $q=10$ and $q=1$, respectively. To plot these curves, one chooses $\theta_0 $ within interval $ [0,\pi/2] $ and imposes a regularity condition on the horizon and shoots to the boundary. Then the quark mass $M$ can be read off from the near boundary behavior of the embeddings. It is clear from this figure that the mass $\theta_0$ is not a single-valued function of $M$. In the following figures, we decided to quote $\theta_0$ values instead of the mass parameter $M$. Since the region where $M$ is not single-valued implies an instability in the system which needs more investigations. Notice that M is dimensionless and the quark mass is defined as  $M_{quark}= M (1/2) T \sqrt{\lambda}$.

As it was proposed in \cite{Kaminski:2009ce}, changing in the sign of $\frac{\delta M}{\delta \theta_0}$ in Figure \ref{Fig1}, implies that the embedding is not stable which happens close to the first order phase transition in the system. More precisely, one may compare it with the van-der-Waals gas, where the unstable region connects two metastable branches as known as overheated and undercooled in the phase diagram \cite{Kaminski:2009ce,Paredes:2008nf}. While the metastable branches are stable against the fluctuations, the unstable branch shows different behavior; a tachyonic mode appears in the QNMs. Interestingly we get the information for this unstable region from Figure \ref{Fig1}. In this figure, we have used the AdS/CFT dictionary and computed the quark mass $M$ from the UV behavior of each D7 brane embedding at fixed $\theta_0$. The vertical dashed lines show explicitly the maximum quark mass. For $q=10, 3, 1$, the embedding angles are $\theta_0=1.12, 1.16, 1.24$, respectively. Interestingly, these values are the same as the $\theta_0$ values in the Figure \ref{Fig4}. One finds that considering the anomalous dimension of the scalar condensate changes the properties of the system. Here the unstable region is affecting by the $q$ parameter. We see that the tachyonic mode shows up for heavior quarks as we are increasing the $q$ parameter. There exist numerical issues for studying the limiting D7 brane embedding at $\theta_0=\pi/2$ so we consider a maximum value of $\theta_0=\pi/2- 0.01$ in our paper.

In this paper, we are interested in fluctuations of the scalar field $\theta$ with mass squared $m^2=-3$. It is shown that the anomalous dimension plays a key role in the analysis of the QNMs. The fluctuations that we consider are space dependent and singlet states on S$^3$ part of the D7 brane world volume. However, in the next section, we turn on an axial field in $\psi$ direction and investigate how it affects the scalar QNMs. We start the calculations by adding a time dependent fluctuation to the embedding with the following ansatz
	\bea \label{anzatsf}
	\Theta (t,z,u)=\theta (u)+\vartheta(t,z,u), \,\,\,\,\,\, \psi=0.
	\eea
		By inserting the above fluctuation ansatz into the DBI action, one gets    
    	\begin{eqnarray}\label{actionfluc}
    	S_{D7} &=& \int dt~du  ~h(u) ~\frac{\cos ^3 \left(\Theta(t,z,u)\right)}{u^5} \nn \\
    	&& \sqrt{1-\frac{u^2}{(1-u^4)}  ~(\partial _t \vartheta(t,z,u))^2+u^2~(1-u^4)~\Theta(t,z,u)^2},\nn \\
        \end{eqnarray}
 
  where $ \theta {(u)} $ is the solution of embedding equation of motion \eqref{embedding}. Assuming plane wave form $ \vartheta(t,z,u)=\text{e}^{-i\left(\omega\, t+k\, z\right)}\delta \theta(u) $ for the fluctuation, and keeping the equation of motion linearized, we find a second order differential equation for the fluctuation. In the following equation, we used \eqref{embedding} to eliminate $\theta''(u)$,
	\begin{eqnarray}\label{fluc1}
		0 &=& \delta \theta ''(u)+A_1(u)~\delta \theta'(u)+A_0(u)~\delta \theta (u),
	\end{eqnarray}	
	where	
	\begin{eqnarray}\label{fluc}
		A_0(u) &=& \frac{1-u^2(-1+u^4)~\theta'(u)^2}{2u^2(-1+u^4)^2}\sec ^2\big(\theta(u)\big)\nn \\
		&& \bigg(-6(-1+u^4)\nn \\
		&& +u^2\left(1+\cos \big(2\theta(u)\right)\big)\left((-1+u^4)~k^2+\omega^2\right) \bigg)\nn \\
		A_1(u) &=&
		\dfrac{1}{u(-1+u^4)(1+u^q)}\big(-3-u^4+(-3+q)u^q\nn \\
		&& -(1+q)u^{4+q}+B(u)~\theta'(u)^2\big)-6\theta'(u)\tan \big(\theta(u)\big)\nn \\
		B(u) &=&
		-12u^2+18u^6-6u^{10}+3(-4+q)u^{2+q}\nn \\
		&& -6(-3+q)u^{6+q}+3(-2+q)u^{10+q}\nn \\\nn
	\end{eqnarray}
	The fluctuation equation \eqref{fluc1} has two independent solutions $ \delta \theta _{\pm} \equiv C_{\pm} (1-u)^{\pm i\omega/4} $. To impose ingoing boundary condition, we set $ C_+=0 $. The fluctuation function is then rescaled as $ \delta \theta (u) \to (1-u)^{-i\omega/4}\tilde{\delta \theta}(u) $ to kill singularities. Using the spectral method, we solve \eqref{fluc} with appropriate boundary conditions. First, we consider the vanishing spatial momentum $k$
		\begin{figure}[tbp]
		\includegraphics[scale=.34 ]{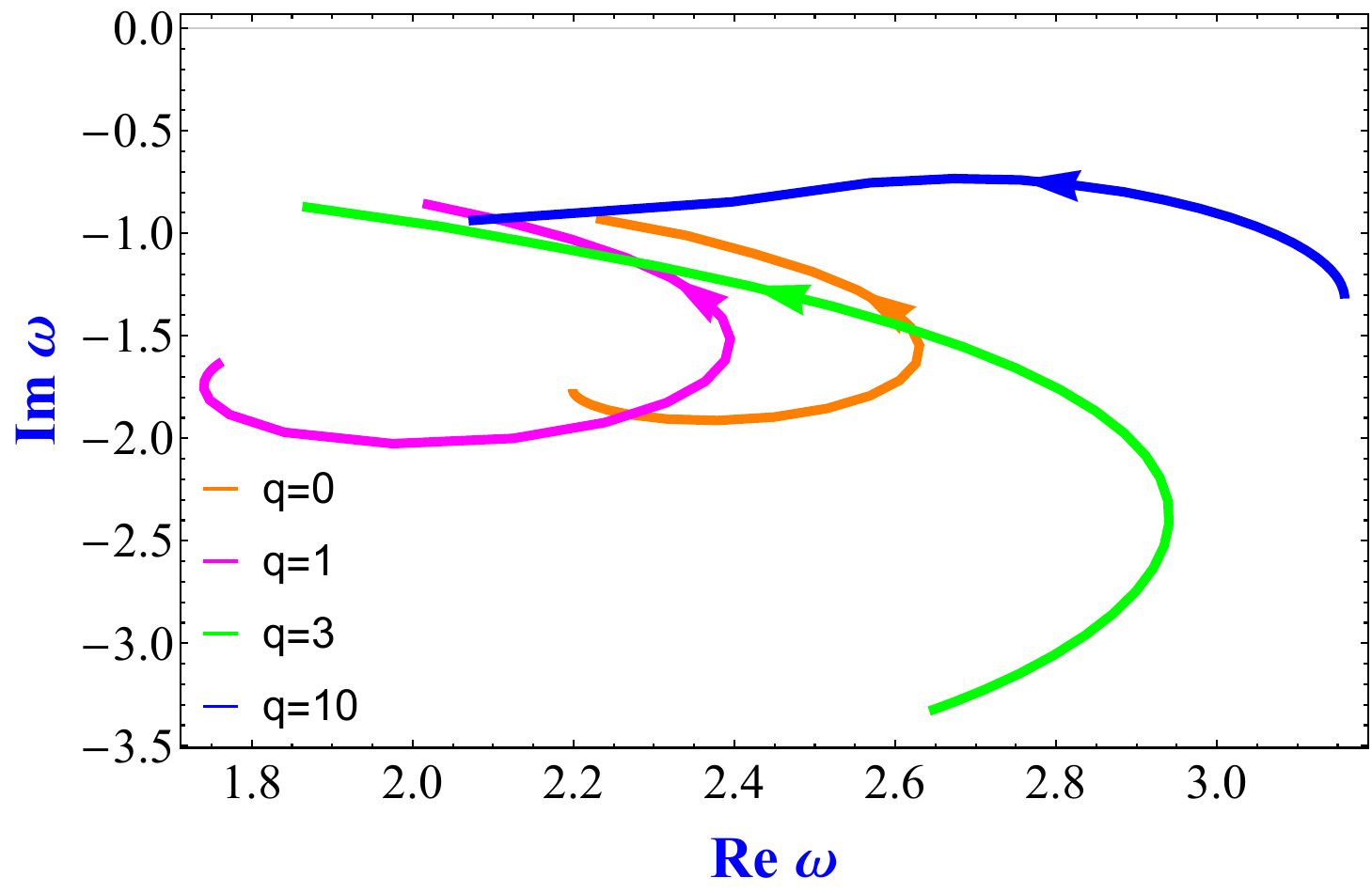}	
		\caption
		{The first QNM in the complex frequency plane at vanishing spatial momentum $k$ as a function of the black hole embeddings using the spectral method. To explain each curve clearly, notice that there are two special points on each curve as starting  $(\theta_0=0)$ and ending $(\theta_0=\pi/2)$ point. At each curve, arrow shows the direction of increasing $\theta_0$ along the curve. For $q=3$ and $q=10$, the massless first QNMs have not been shown here because they are pure imaginary as one finds in Figure \ref{Fig2}. Considering the starting point in each curve, from left to right $q=1,0,3$ and $10$, receptively. }\label{Fig3}
	\end{figure}     
   	and consider massless quarks $\theta_0=0$ corresponding to flat D7 brane embeddings. To do the calculations, we fix the temperature and vary the $q$ parameter. One should notice that in the dual theory, the meson spectrum is continuous and quarks are deconfined.	Figure \ref{Fig2} shows the imaginary part of the first ten quasinormal mode frequencies as a function of the real part. In Table 1, we give the numerical values for $q=0$ case. In this paper, all numeric results for QNMs computed using 61 points in the grid which was judged by a comparison with a grid with 81 points as Table 1. 
   	
   In Figure \ref{Fig2}, one finds that by increasing $q$ a massless pure imaginary mode appears in the spectrum. We checked carefully this new observation for different values of $q$ and found that there exists a critical value for $q_c$ in which for a larger value of $q$ such modes appear in the system. In the case of $q=0$, by using the shooting method, these modes were not found in \cite{Hoyos-Badajoz:2006dzi}. However, by changing the numeric method to the relaxation method, authors of \cite{Kaminski:2009ce} found pure imaginary modes for massless modes. We did not observe these modes by using the spectral method, too. It would be interesting to compare these different methods and investigate which method gives the correct result. It is beyond the scope of this study. 
   	
   As it was expressed, the parameter $q$ is related to a finite value of $\frac{N_f}{N_c}$ then we are studying a QCD-like gauge theory at strong coupling and finite temperature. The main idea is that for a given $x$ there is a choice of $q$ that best matches expected QCD physics. However a precise relation is not known because of the bottom-up nature of the model. It is a very interesting result that by increasing $ q $, a pure imaginary appears in massless modes for $ q_c \approx 2.70319 $, actually a tower of these damping modes shows up for $ q>q_c $ as we see in the bottom-right plot of Figure \ref{Fig2}. In Figure \ref{Fig2-2}, we study how the first QNM is colliding at the imaginary axis. In this figure we plotted the real part of the first QNM as a function of $q$. One finds that it vanishes at $ q=q_c $ and the mode becomes a pure imaginary one. As it was discussed in \cite{Kaminski:2009ce}, the system would be unstable in this regime which is close to the first order phase transition.

	\begin{figure}[tbp]
		\includegraphics[scale=.28]{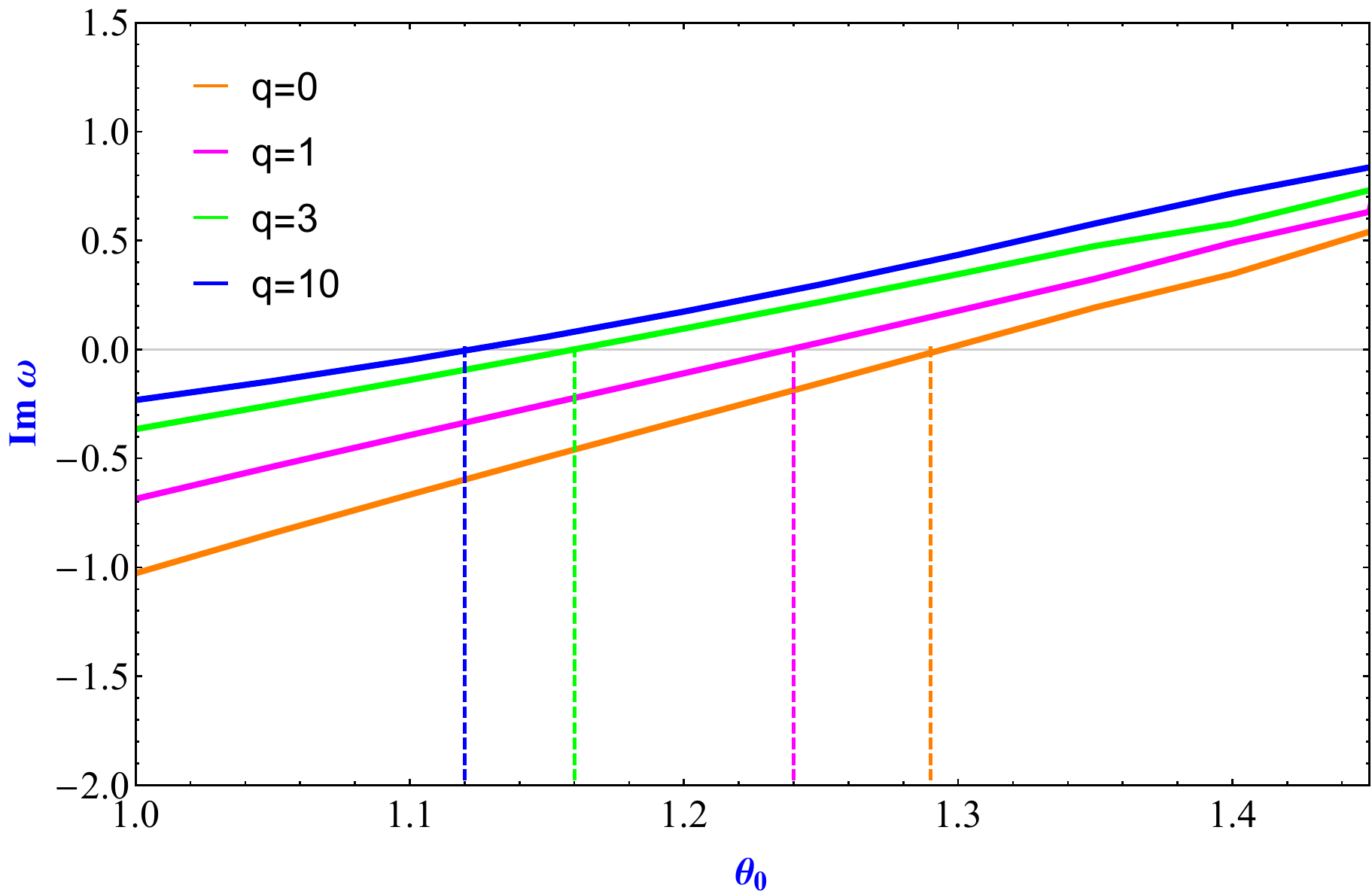}		
		\caption
		{Pure imaginary QNMs at vanishing momentum as a function of the quark mass.  An instability occurs by changing the sign of the pure imaginary mode. From bottom to top $q=0,1,3,10$. The dashed vertical lines show the quark mass values where the instability occurs in the system. Interestingly, they match with Figure 1. }\label{Fig4}
	\end{figure}
	
	Next, we consider the first quasinormal frequencies of massive quarks in Figure \ref{Fig3}. This figure includes curves for different non-zero $q=1,0,3,10$ from left to right. We describe the curve for $q=0$, first. Zero quark mass $(\theta_0=0)$ corresponds to the point with the smallest value of the real part and by increasing the mass along the curve, the real part reaches a maximum value and then decreases to reach the endpoint. Also, the imaginary part turns by increasing $ \theta_0 $. Now we see in this figure how $q$ changes the general feature of the $q=0$ curve. For $q=3$, the general shape does not change but it moves slightly to right-hand side of the $q=0$ curve, while the $ q=1 $ curve moves left. But $q=3$ increases both the real part of complex frequency values, significantly and the imaginary turning point disappears. For $q=10$, the turning behavior along both the imaginary and real parts disappears. In contrast with the $ q=3 $ case, the absolute values of the imaginary part become smaller for $ q=10 $.  One finds that the starting point (small mass) has been affected stronger than the endpoint (large mass) of each curve. 
	
	Along each curve in Figure \ref{Fig3}, one finds that the corresponding mode becomes more and more stable. However, as we have seen in Figure \ref{Fig1} a tachyonic mode appears in the system which depends on the $q$ parameter or the anomalous dimension of the condensate. Still, we are staying in the region close to the first order phase transition and keeping the black hole embeddings. One should notice that the endpoint of the trajectories in Figure \ref{Fig3} corresponds to the limiting D7 brane black hole embedding. This is the one that is touching the black hole horizon. Because of the tachyonic mode in the spectrum, it is not expected to reach this limiting embedding. 
	
 It is an interesting observation that the effect of $q$ on the QNMs depends on the masses of quarks. Generally, the ordering of embeddings and QNMs for an arbitrary value of $ \theta_0 $ strongly depends on the value of $ q $ whether $ q<q_c $ or $ q>q_c $. Perhaps it implies an interesting physics close to the critical value $ q_c $.  
		\begin{figure}[tbp]
		\includegraphics[scale=.28]{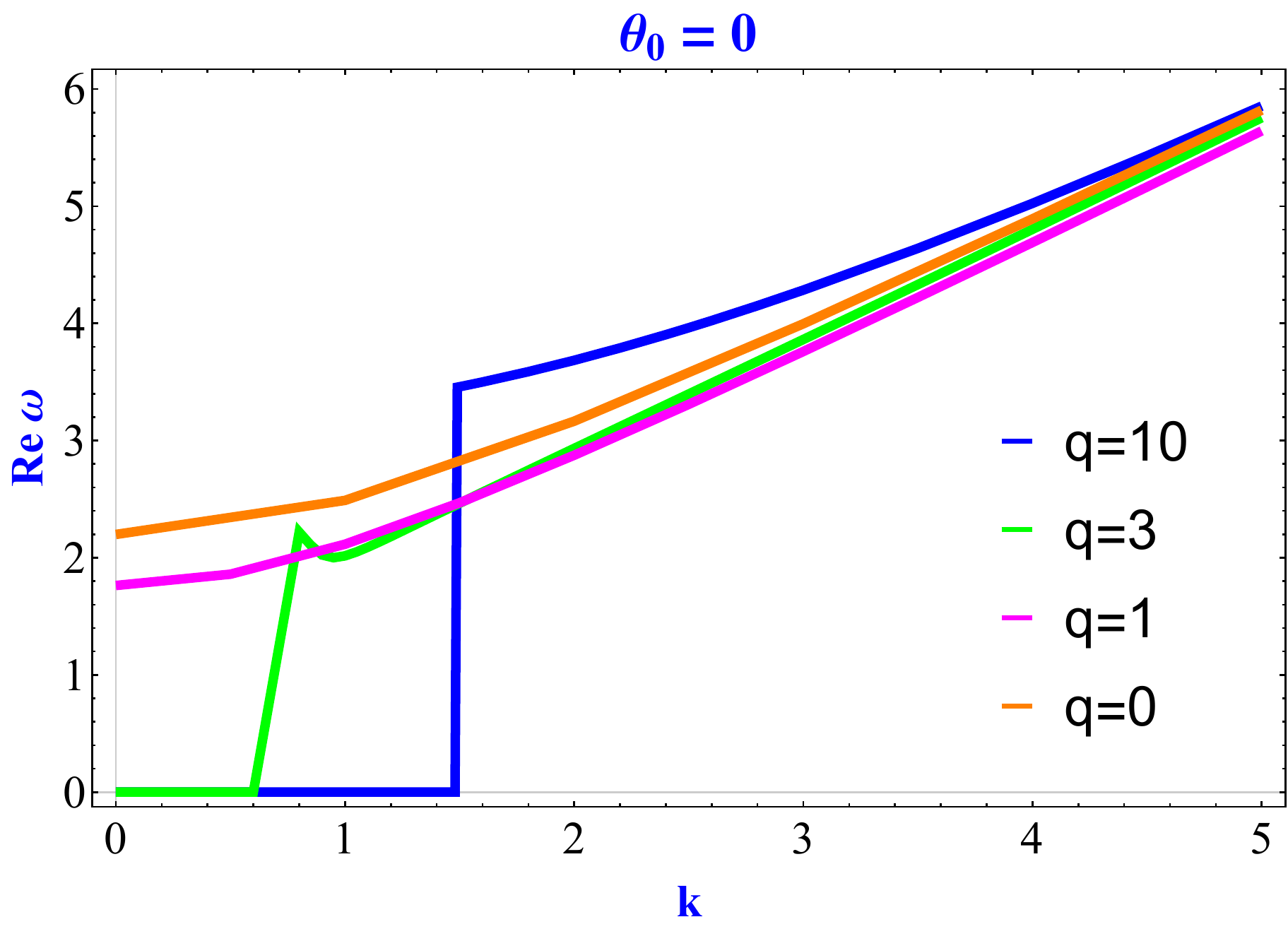}
			\includegraphics[scale=.282]{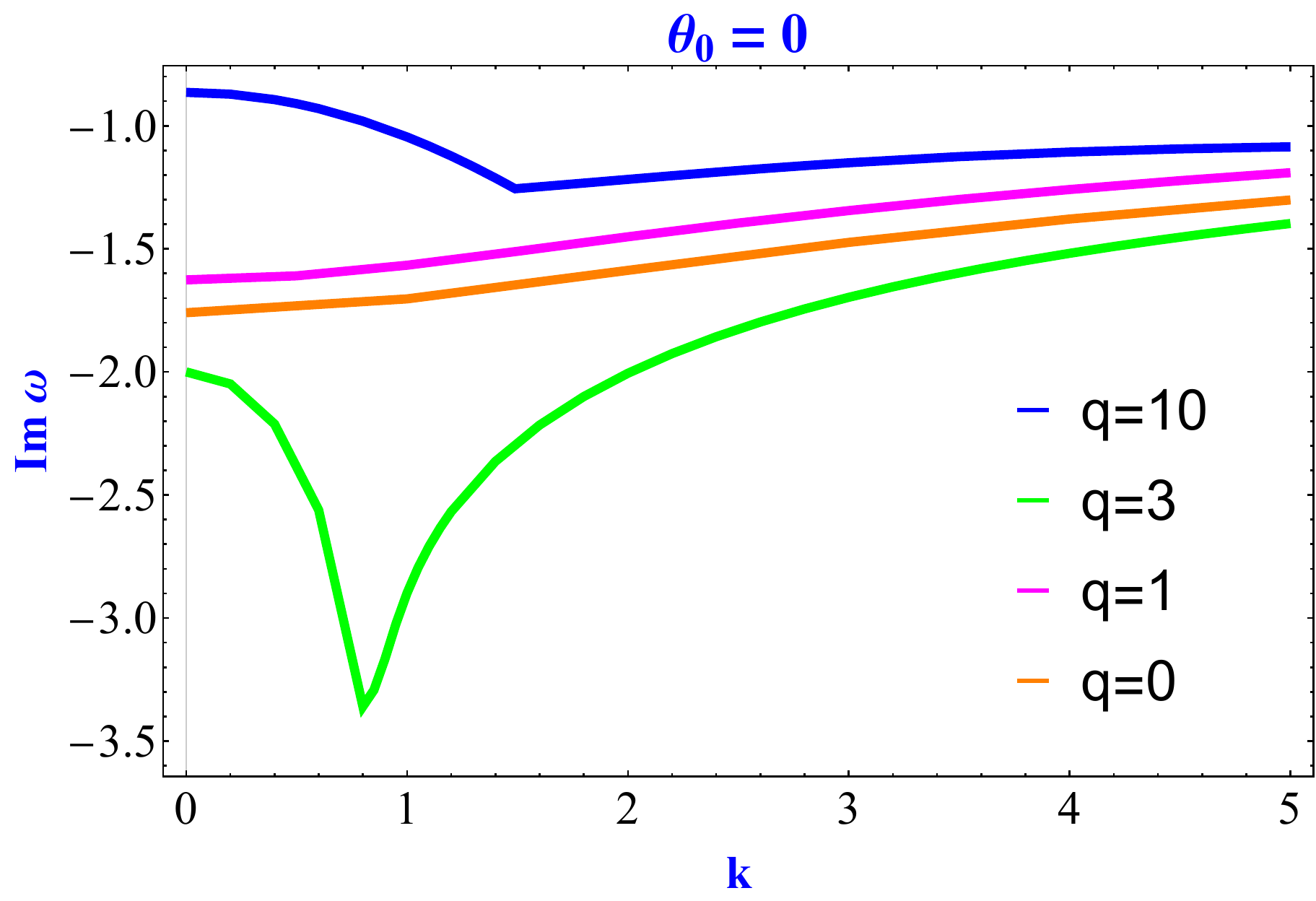}	
		\caption
		{ Dispersion relation for the first QNM for massless quarks $(\theta_0=0)$. As Figure 2, for $q>2.705$ the real part of the mode vanishes at small momentum.	}\label{Fig5}
	\end{figure}

Figure \ref{Fig4} shows the behavior of the pure imaginary modes as a function of $ \theta_0 $ for different values of $q=0,1,3, 10$ from bottom to top. As it was explained, there is no pure imaginary mode for massless ($\theta_0=0$) quarks for $q<q_c$ like $q=0$ and $q=2$ in Figure \ref{Fig2}. Interestingly, the vertical lines in Figure \ref{Fig1} implies to instability in the system. Indeed in Figure \ref{Fig4}, it is clear that there exists a critical value for the $\theta_0$ in which the scalar QNM becomes tachyonic and makes the system unstable. Because the pure imaginary mode changes the sign from negative to positive. Before this value of $\theta_0$ is reached, it is expected that the system will undergo a phase transition to Minkowski solutions. At $q=10, 3, 1, 0$, the corresponding $\theta_0$ values are $1.12, 1.16, 1.24, 1.29 $, respectively. Those values are the same as the Figure \ref{Fig1}.

\begin{figure}[tbp]
	\includegraphics[scale=.45]{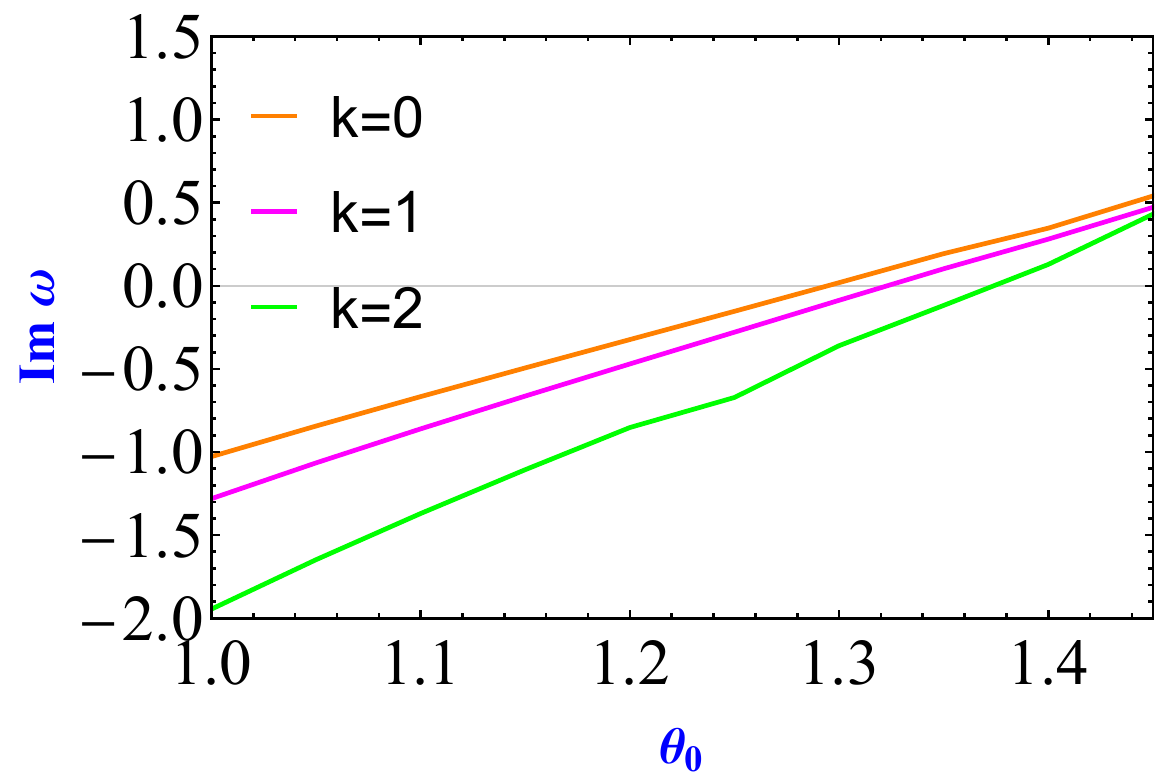}	
	\caption
	{Effect of momentum on the pure imaginary QNM. From top to bottom $k=0,1,2$.	}\label{Fig6-1}
\end{figure}
Our study helps to understand better the behavior of the scalar massless QNM and confirms that it becomes tachyonic for massive quarks. So that the D7 brane embedding is not stable and one should find the true ground state of the system \cite{Kaminski:2009ce}. We showed that considering the anomalous dimension leads to the appearance of the pure imaginary mode in the system. Comparing with the other studies, one finds that a new instability in AdS space-time with a scalar potential was found in \cite{Gursoy:2016ggq}. Also thermodynamics of Einstein-scalar field theories shows a rich structure \cite{Gursoy:2018umf}. In general, dynamics near a first-order phase transition have been studied from holography in \cite{Bellantuono:2019wbn}. In the D3/D7 configuration, the emergence of a positive pure imaginary quasinormal mode signifies system instability. This is attributed to the fact that a QNM with a positive imaginary component implies an exponentially increasing perturbation over time, rather than a diminishing one. The D3/D7 framework comprises D3-branes, which form the foundation of AdS space, and D7-branes, which infuse the gauge theory with flavor variables like quarks. The system's dynamics, especially at finite temperatures, are often analyzed through the QNMs of the branes, reflecting the system's reaction to minor disturbances. The transition of a pure imaginary mode to a positive value may arise when the quark mass, introduced via the D7-branes, surpasses a specific limit. This event can cause the scalar fields on the D7-branes to become tachyonic, indicating a negative mass squared and signaling inherent instability. Consequently, the system endeavors to find a new, stable ground state. This instability is linked to a phase transition within the gauge theory aspect of the AdS/CFT duality, suggesting that the initial D7-brane arrangement is not the genuine ground state and that the system will evolve into an alternative configuration that reduces free energy.
	
One should notice that the presence of a positive pure imaginary mode in the QNM spectrum is widely recognized as an indicator of an unstable phase across diverse physical systems. Investigating these instabilities is vital for comprehending the thermodynamics and transport characteristics of the strongly coupled plasma and for probing the phase diagram of the related gauge theory.

We presented the first QNM at vanishing spatial momentum $k$ in Figure \ref{Fig2}. Now we study the effect of non-zero $k$ by solving \eqref{fluc}. We checked numerically that by increasing quark mass the real part shows a turning behavior similar to Figure \ref{Fig3}.  We investigated the effect of the momentum of modes and found that by increasing $k$ the curves in the complex plane move to the right. As an important effect, we saw that $q_c$ grows by increasing $k$. Interestingly, the pure imaginary mode for massless quarks depends on the value of $k$ as well, we conclude that there is a critical value for momentum $k_c$ in which the pure imaginary mode disappears for $ k>k_c $ at fixed $q$. For example, at $ q=3 $, we found that $ k_c\approx1 $. 
 
 We show the dispersion relation for the first QNM for massless quarks in Figure \ref{Fig5}. The parameter $q$ is changing from $q=0$ to $q=1,3, 10$ in this figure, we present the results for momenta between $0$ and $5$. By increasing spatial momentum $k$, the real part of the frequency increases and the mode gets more energy but at small $k$ one finds different behavior for $q$ larger than $q_c$, i.e $q=3$ and $q=10$. As we have explained, in these cases the real part is zero.  This is the expected behavior that we learned from Figure \ref{Fig2} where for larger $q$ a pure imaginary mode appears in the spectrum. However, we see that even at finite momentum the real part of the first QNMs is zero but at $k_c$ the real part start growing. We observed that the real part of the QNMs exhibits a spike near $ k_c $, as illustrated in Figure 6. As we increase $ q $, the spike diminishes, which is evident at $ q=10 $ in this figure. Such singular behavior of the dispersion relation needs more investigations. That would be interesting to do the WKB study of the Schrodinger potential to get some intuition about these QNMs. The imaginary part of the mode in Figure \ref{Fig5} confirms this behavior and describes the $q=3$ curve clearly. Modes with higher values of $k$ do not depend significantly on $q$ and move towards the $q=0$ curve. 
 
 The behavior of the imaginary part of the frequency, denoted as $ Im\omega $, is indicative of the stability of the system. When $ Im\omega $ is positive, it suggests the presence of an instability in the system. In the context of the D7 brane fluctuation within the dual QCD-like gauge theory, this instability is associated with a particular value of mass, referred to as the critical mass.	As the momentum of the fluctuation increases, the critical mass at which the instability occurs also rises. This relationship can be visualized in Figure \ref{Fig6-1}, where the dependency of the critical mass on the momentum is plotted. One finds that a larger quark mass is required to reach the point of instability. That is an imprtant observation. Because the increase in critical mass with momentum could be related to the energy required to excite certain modes of the brane. These modes, when sufficiently energized, could lead to the observed instability. Furthermore, exploring the critical mass's dependence on momentum could provide insights into the phase structure of the dual gauge theory. It might reveal how the theory behaves near the phase transition points, where the properties of the system undergo significant changes.
\section{Adding pseudoscalar field}
We now switch on the pseudoscalar field $\psi$ on the world volume of the D7 brane. In the dual bulk gravity, we fluctuate the embedding D7 branes as 
\bea
\Theta=\theta+ \vartheta, \,\,\,\,\, \psi= \psi_0.
\eea 

	\begin{figure}[tbp]
	\includegraphics[scale=.55]{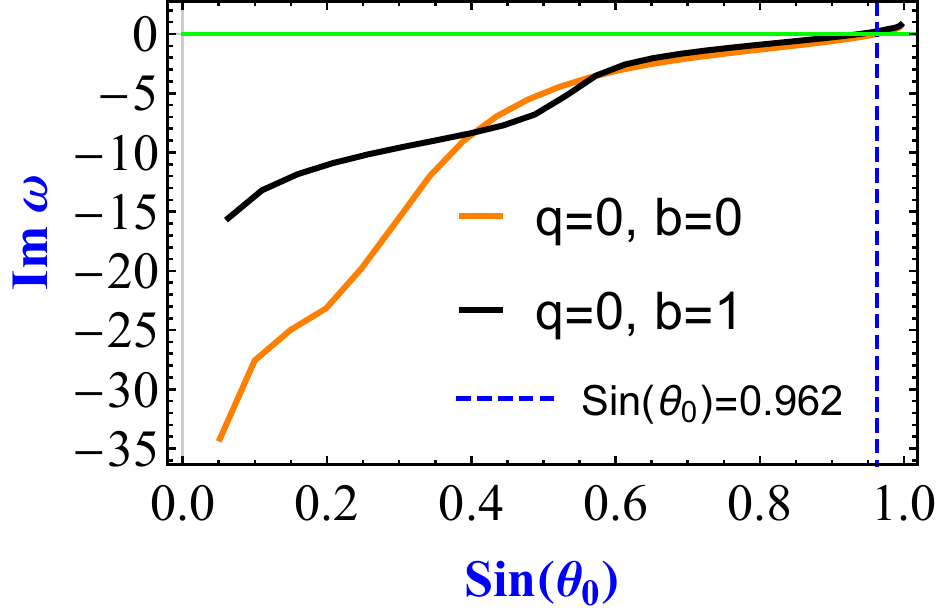}
	
	\caption
	{Effect of pseudoscalar field $b$ on the first pure imaginary quasi normal mode at $q=0$. The vertical dashed line shows the crossing point with the real axis where an unstable mode appears in the system. Its value is the same as \cite{Kaminski:2009ce}. }\label{Fig-8}
\end{figure} 
	\begin{figure}[tbp]
	\includegraphics[scale=.45]{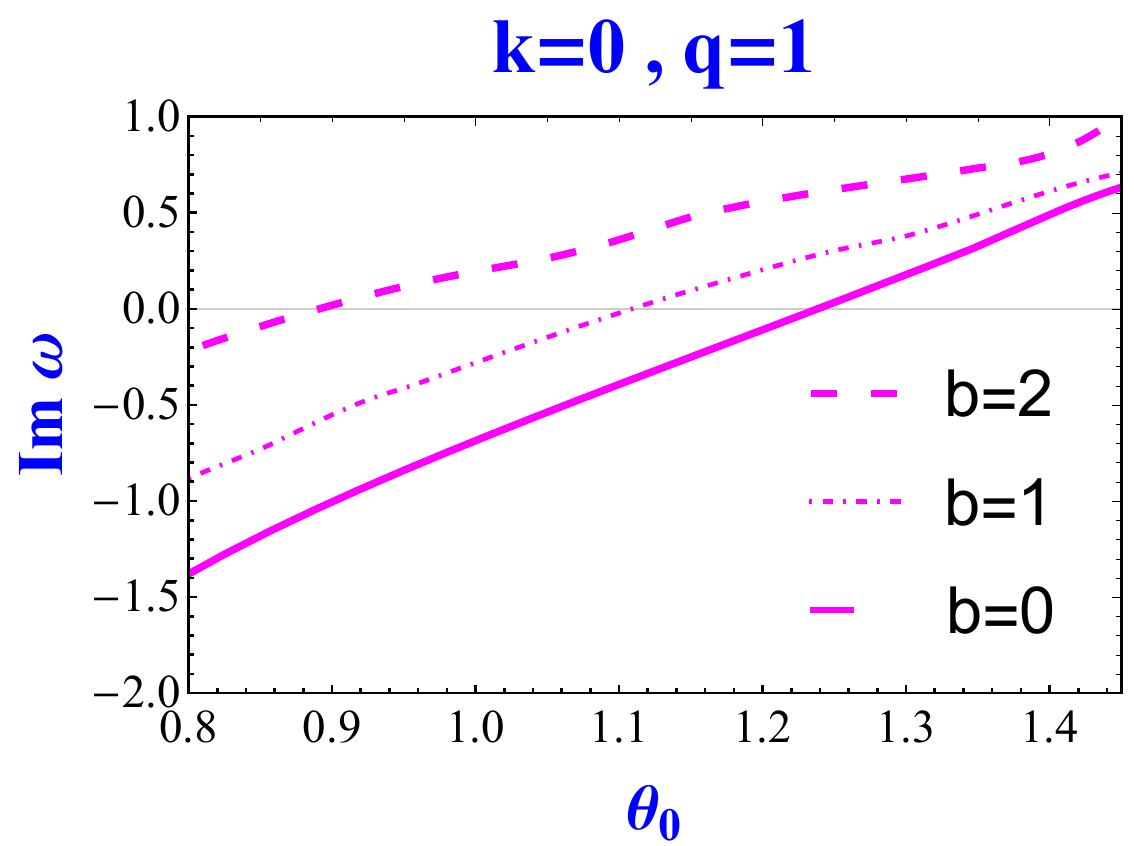}	
	\caption
	{Pure imaginary QNM in the presence of pseudoscalar field $b$ at fixed value of $q=1$. From top to bottom $b=2,1,0$.  	}\label{Fig-9}
\end{figure}
\begin{figure}[tbp]
	\includegraphics[scale=.25]{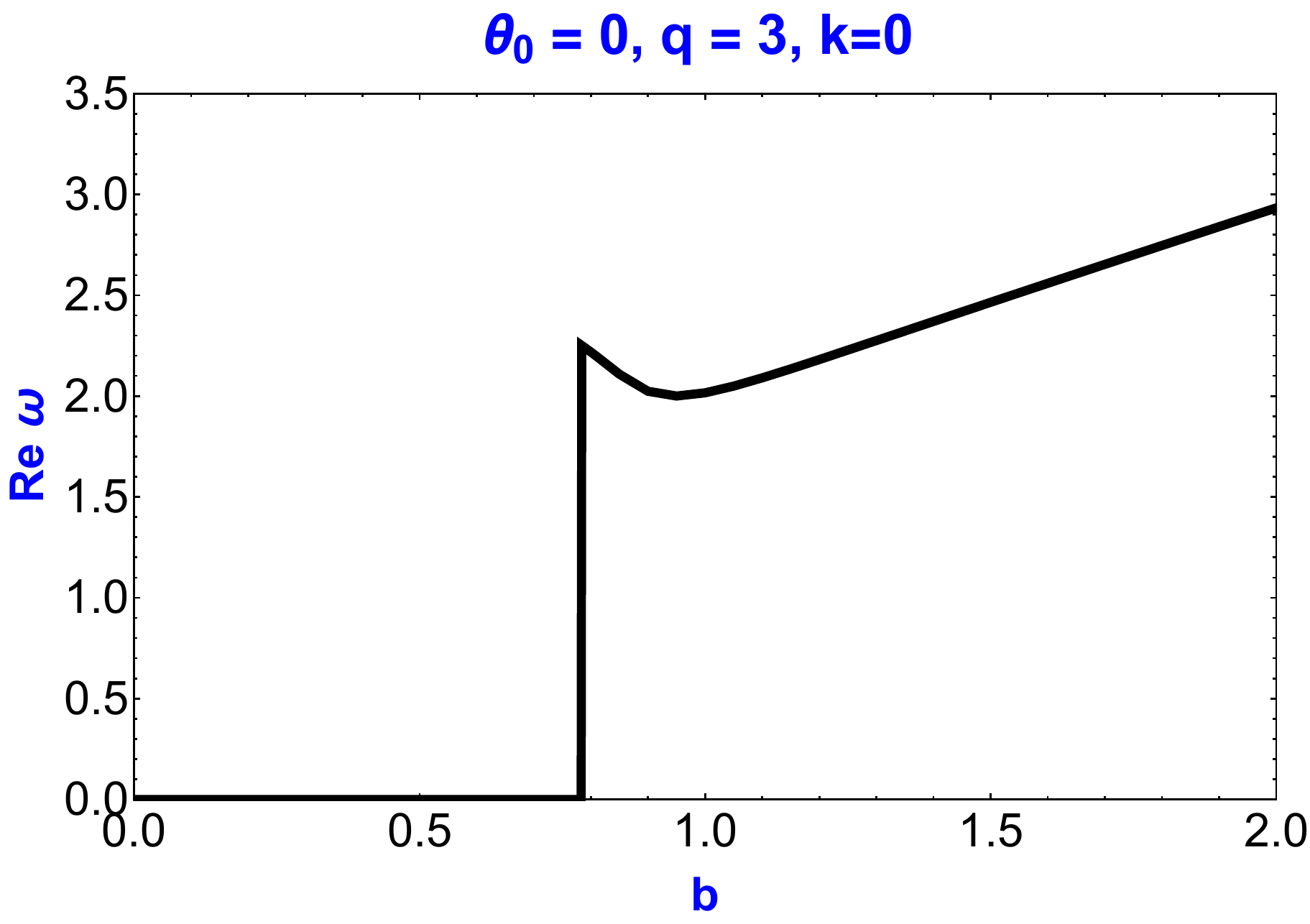}
	\includegraphics[scale=.25]{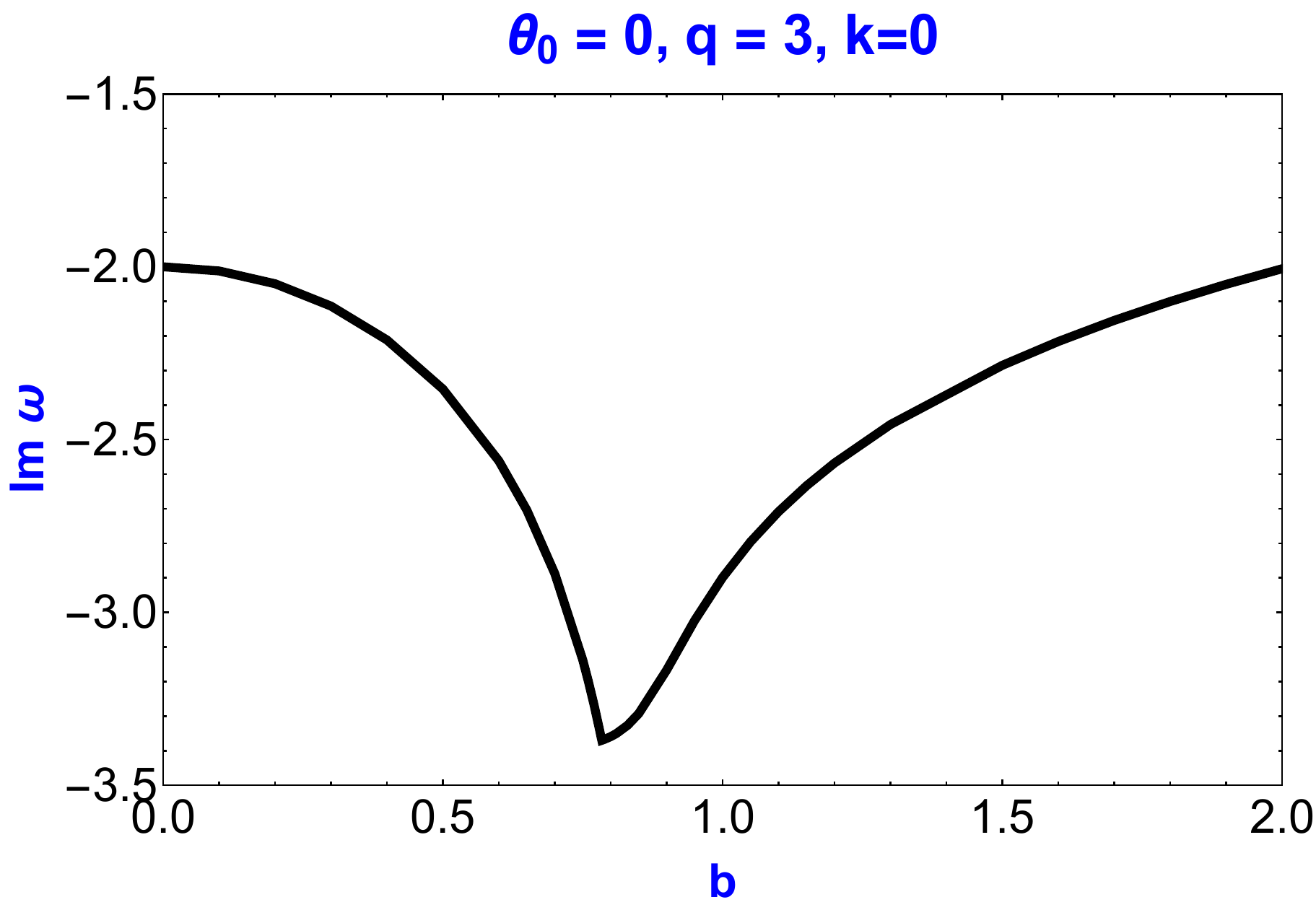}
	\caption
	{Effect of pseudoscalar field $b$ on the real and imaginary parts of the QNMs at fixed $q=3$. The value of $q$ is larger than $q_c$ and we consider zero quark mass $(\theta_0=0)$. }\label{Fig-10}
\end{figure}

 We checked that the equations of motion corresponding to the fluctuation of scalar and pseudoscalar decouple from each other like \cite{Bu:2018trt}. Here, we ignore the fluctuation of the pseudoscalar similar to \cite{Ishigaki:2021vyv}, where they perturbed the gauge and scalar fields. Also in \cite{Ishigaki:2020vtr} fluctuations on pseudoscalar and gauge fields have been studied. Its interpretation from the AdS/CFT correspondence has been studied in \cite{Myers:2007we}. One finds that near the boundary, on the world volume of the D7 brane embedding, the leading term in the expansion of this field determines the phase of the quark mass while the subleading term determines the expectation value of the related boundary operator. 
 
 We use the same simple ansatz as \cite{BitaghsirFadafan:2020lkh}, which introduces the phase of quark mass as the following
 \bea
 \psi_0=b z
 \label{psi}
 \eea
 in which $b$ is an axial field and $z$ is one of the boundary field theory space directions. As \cite{BitaghsirFadafan:2020lkh}, a more general ansatz than equation \eqref{psi}, in which the pseudoscalar can also have non-trivial dependence on the holographic radial direction $u$, would be $\psi = b\,z + \psi(u)$. In addition of the new term proportional to $\lambda(\rho)$ in the DBI action \eqref{dbi1}, the argument of \cite{BitaghsirFadafan:2020lkh} is unchanged and $\psi$ should not depend on $u$. In the presence of a magnetic field, a similar ansatz has been considered in \cite{Kharzeev:2011rw, Bu:2018trt} and the chiral helix studied. As it was studied in \cite{BitaghsirFadafan:2020lkh}, in the presence of an electric field, the $b$ field causes interesting transport due to the axial anomaly. In this paper, we take this ansatz and expect that our study could be related to such phenomena when the anomalous dimension of quark condensate has been added. Comparing with \cite{BitaghsirFadafan:2020lkh}, one does not expect a background axial gauge field in QCD. But because of a strong magnetic field in the core of neutron stars, the hot and dense QCD phase diagram will be changed significantly so that new inhomogeneous phases appear. It was shown that the energy spectrum of quasi-particles in this phase is the same as the WSMs \cite{Tatsumi:2018ifx}. Also it was argued that the magnetic dual chiral density wave is a candidate for the quark matter phase in the core of neutron stars \cite{Ferrer:2021mpq}.

We extend our analysis from the previous section, adhering to the same methodologies and assumptions. In the field theory side the free parameters are $M$, $b$ and $T$, all with dimensions of mass. We will plot the physical quantities in figures of this section in units of $T$. The fluctuations equation at finite momentum is given in Appendix \ref{B}, too. First we consider vanishing momentum and study the first QNMs in the system. Figure \ref{Fig-8} shows effect of pseudoscalar field $b$ on the first pure imaginary quasi normal mode at $q=0$. The vertical dashed line shows the crossing point with the real axis where an unstable mode appears in the system. To check validity of our numerics, we presented the horizontal axis in terms of $Sin(\theta_0)$. One finds that its value is the same as \cite{Kaminski:2009ce}.

Next, we choose $q=1$ and present the pure imaginary modes in Figure \ref{Fig-9}. We see that the tachyonic mode is present in the system where it signals an instability. The introduction of the axial field $b$ influences the critical mass at which this tachyonic mode appears. Notably, the critical mass decreases as the axial field is turned on, suggesting that the axial field has a stabilizing effect on the system.  We found the same structure at different values of $q$. We also found that there is no pure imaginary QNM for massless quarks at $q=0$ but non zero $b$. This behavior is intriguing because it contrasts with the effect of momentum on the critical mass, where an increase in momentum leads to an increase in the critical mass required for the onset of instability. The axial field $ b $, however, induces the opposite effect, reducing the critical mass threshold for instability. Further research should focus on constructing a detailed phase diagram that maps out the regions of stability and instability as functions of both momentum and the axial field $ b$. Such a diagram would provide a more comprehensive understanding of the conditions under which the system transitions from a stable to an unstable state. Additionally, it would be beneficial to explore the underlying mechanisms that govern the influence of $b $ on the system, potentially revealing new insights into the nature of tachyonic instabilities.

In Figure \ref{Fig-10}, we illustrate the influence of the pseudoscalar field $b$ on the real and imaginary components of the QNMs while maintaining $ q=3$. Notably, in this case, parameter $q$ exceeds the critical value $q_c$. We also examine the case of a massless quark. It is observed that there exists a critical pseudoscalar field value $b_c$, beyond which the massless pure imaginary mode ceases to exist for $b > b_c$. Our analysis confirms that for $q=3$, the critical value is determined to be $ b_c=0.75$. One finds in this figure the spike behavior close to $b_c$.

\begin{figure}[tbp]
	\includegraphics[scale=.28]{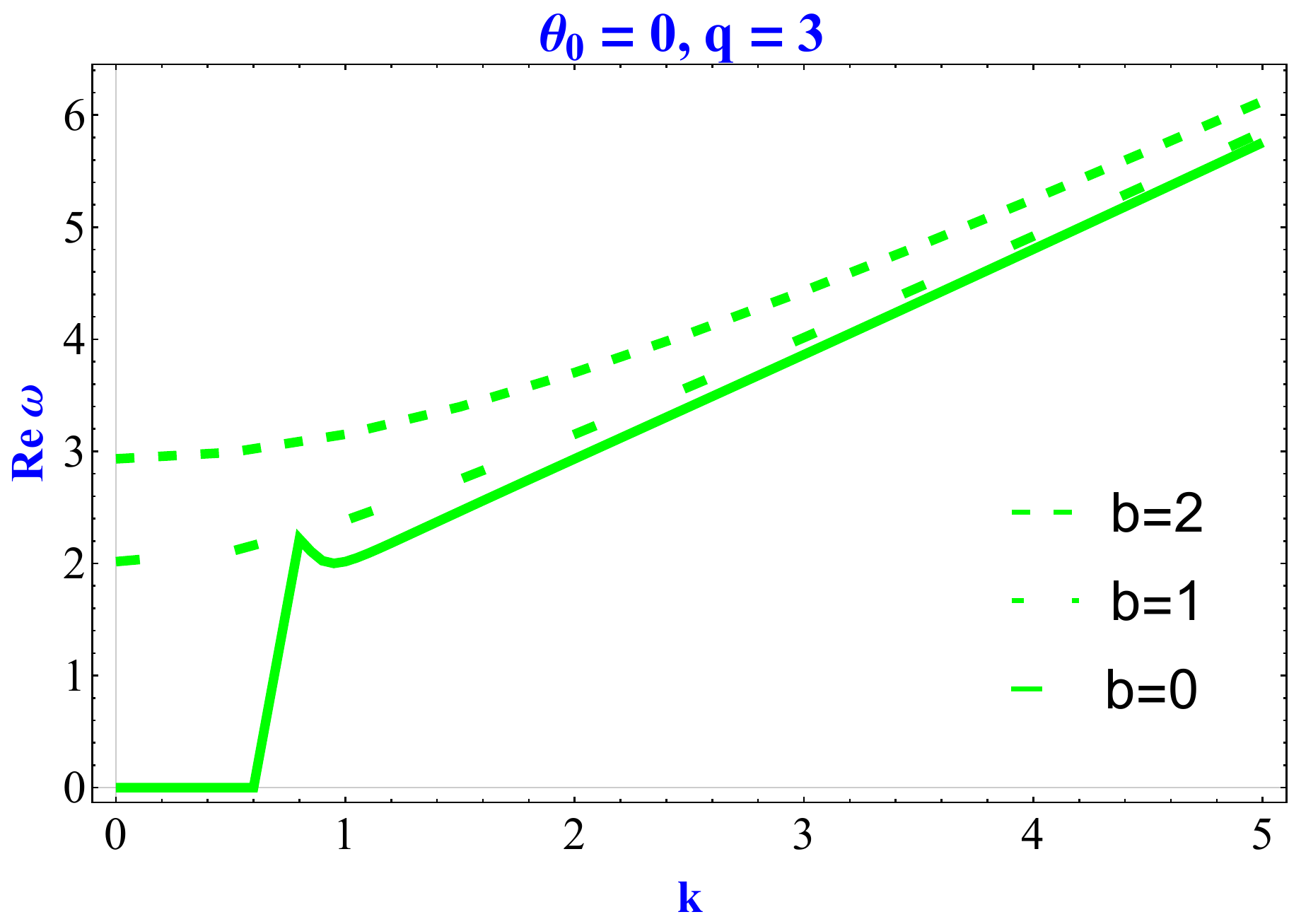}	
	\includegraphics[scale=.28]{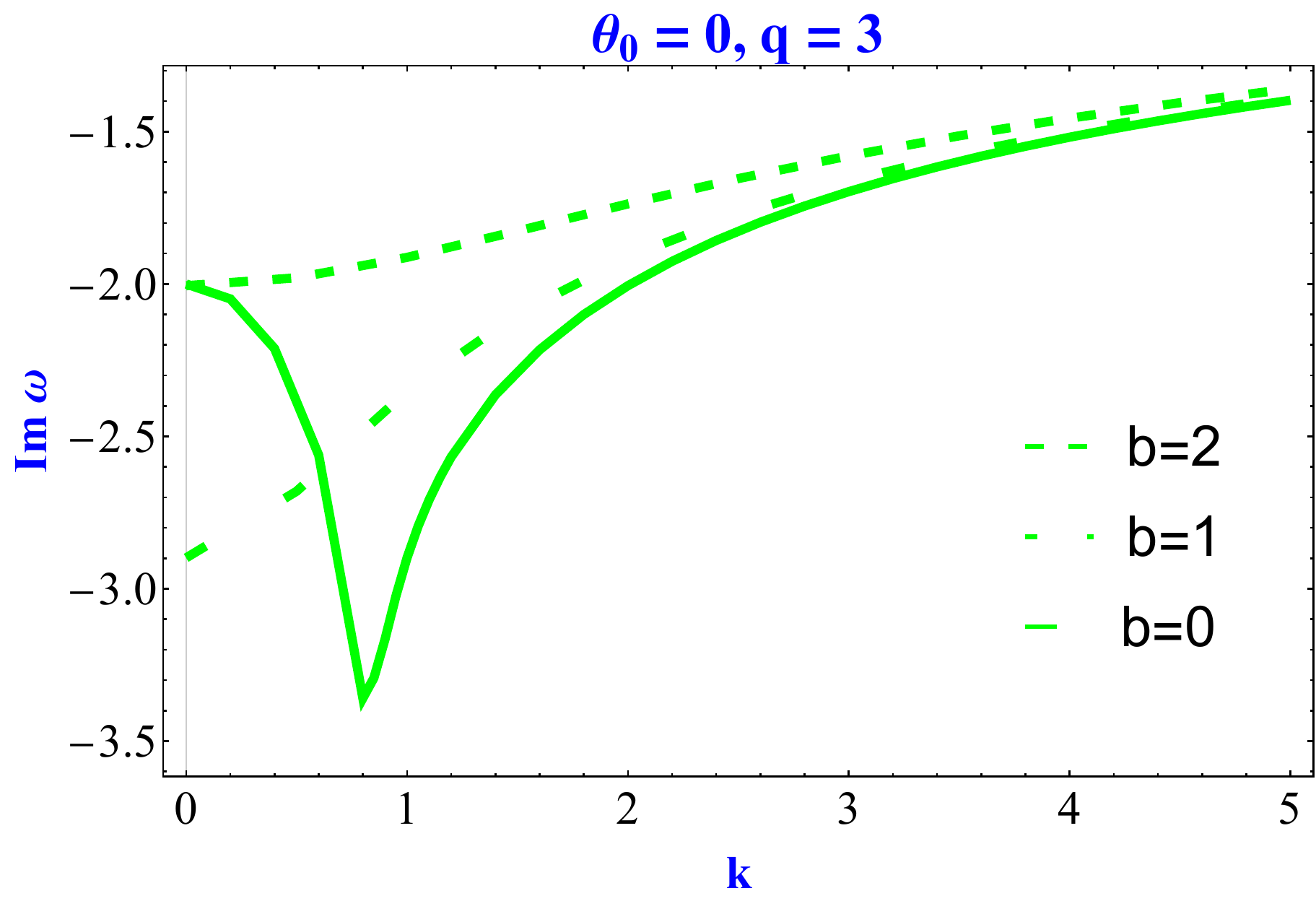}
	\caption
	{Effect of pseudoscalar field $b$ on the dispersion relation at fixed $q=3$. The value of $q$ is larger than $q_c$ and we consider zero quark mass $(\theta_0=0)$. From top to bottom $b=2,1,0$.}\label{Fig11}
\end{figure}

The dispersion relation depicted in Figure \ref{Fig11} elucidates the behavior of the first QNMs for massless quarks at $q=3$, with the momentum $k$ varying from $ k=0 $ to $ k=5$. The influence of the axial field $b$ on the QNMs is more pronounced at lower values of $ k$, affecting both the real and imaginary parts of the frequencies. As $k$ increases, this effect gradually diminishes, indicating a reduced sensitivity of the QNMs to the axial field at higher momenta. For massless quarks, the real part of the frequency initially increases with $k$, reflecting the energy imparted to the quarks by the axial field. As $k$ grows further, the real part of the frequency increase monotonically which does not depend on the value of $b$ significantly. Massive quarks exhibit a similar trend in the real and imaginary parts of the frequencies with varying $k$, but the overall magnitude of the frequencies is affected by the quark mass. These observations suggest that the axial field $ b $ plays a crucial role in determining the dynamics of the QNMs, particularly in the low-momentum regime. The results for massless and massive quarks provide valuable insights into the interplay between the axial field and the quark momentum, offering a window into the fundamental properties of the system under study.

	\section{Summary}
In this paper, we have studied an adapted D3/D7 brane setup that included a running anomalous dimension, $\gamma$, for the quark condensate. This allows us to describe the chiral restoration phase transition away from the phase of deconfined massive quarks. A nontrivial $h$ function has been considered as \eqref{profileh1} with a free parameter $q$ that specifies the chiral symmetry breaking and anomalous dimension behavior in the theory. We explored the higher values of $q$ corresponding to the walking theories where the coupling constant changes slowly over a long range of energy scales. The dual gauge theory is a QCD-like theory at finite $N_f/Nc$.

We explained there are two different D7 brane embeddings, Minkoski and black hole embeddings and a first order phase transition occurs between them. We have studied fluctuations of the scalar sector on the D7 probe brane and found a tachyonic mode close to the first order phase transition. It was shown that the purely imaginary mode crosses the real axis and therefore makes the system to become unstable at specific qaurk mass values. An excellent numerical agreement found between this crossing point and where the maximum mass of quark is reaching, see Figures \ref{Fig1} and \ref{Fig4}. At zero and finite momentum, we determined the first quasinormal frequency of the scalar sector. We calculated the QNM trajectories in the complex plane. In general, they move along a curve where initially the imaginary and real parts can grow but then move continuously towards smaller values. The endpoint of this curve is the D7 probe brane embedding that is touching the black hole horizon corresponding to $\theta_0=\pi/2.$ However before reaching on it, we show that the embeddings become unstable because of the tachyonic mode in the system. It was shown that how the $q$ parameter is affecting this phenomena. As a conclusion when $q$ increases the quasinormal modes start colliding in the imaginary axis and moving down in the complex frequency plane. The number of purely QNMs depend on the value $q$ parameter. The critical $q_c$ depends on the mass of the quarks, for massless quarks  $q_c \approx 2.70319$. 

The scalar flavor fields are $\theta$ and $\psi$ that describe transverse directions to the D7 brane embedding. In the dual picture, the scalar field $\theta$ is dual to the quark condensate and the phase transition occurs when the Breitenlohner Freedman (BF) bound is violated by the scalar field. For black hole D7 brane embedding, we studied the fluctuations of $\theta$ field at vanishing and finite spatial momentum, $k$. Our analysis shows that there is a critical $q_c$ about $2.70319$ for massless quarks so that a new pure imaginary mode appears in the complex frequency plane. As an important observation, we found that the existence of such a mode depends on the value of momentum $k$ in the theory. 

A key factor in the stability of QCD is the mass of the quarks, which serve as the system's carriers. Our research indicates that the quark mass substantially influences the stability of different modes within the QCD plasma. Specifically, the growth rate of instabilities, denoted by the imaginary component of the frequency, is affected by the quark mass. Additionally, carrier mass is vital during the thermalization phase post high-energy heavy-ion collisions. As the QCD plasma progresses, instabilities that depend on mass aid in the isotropization and stabilization of the plasma. Comprehending these instabilities is crucial for analyzing data from particle colliders and for developing precise models of the early universe's QGP. Although our analysis pertains to a dual gauge theory akin to QCD, it's important to recognize that in theoretical models, the carrier mass is not just a static parameter but a dynamic element that can govern system stability. The relationship between carrier mass and instability provides profound insight into the strong force's behavior within the framework.
 
We can summarize the main results as follow:
\begin{itemize}
	\item There is a critical mass for quarks in which the sign of the pure imaginary frequency changes and makes the system unstable.
	
	\item The critical mass is impressed by parameter $ q $ and momentum $ k $ so that it becomes smaller by increasing $ q $ and gets larger by raising $ k $.
	
\end{itemize}

The other main result of this paper was the effect of the pseudoscalar field $\psi$ on the QNMs of the scalar field. We take the simple ansatz $ \psi=bz $ of \cite{BitaghsirFadafan:2020lkh} where $b$ is an axial field and $z$ is one of the spatial boundary coordinates. In the presence of an electric field, one finds anomalous transport related to Weyl semimetals \cite{BitaghsirFadafan:2020lkh}. We found that at small quark mass, the effect of the $b$ field is stronger than the massive ones. It is shown that $q_c$ increases also by increasing $b$ which is interesting from a phenomenological point of view. At $q=0$, adding $b$ could be related to topological phenomena like Weyl semimetal phases in condensed matter, but in QCD quarks have color degrees of freedom which make the phase space of the system much richer. It would also be interesting to generalize this study at finite density for scalar or vector fields and study the vector fluctuations in the system to investigate the crossover behavior from the hydro regime to the quasiparticle one. It is also interesting to investigate the phase diagram of the QCD-like theory and study how the critical temperature depends on $q$ parameter.
	\section*{Acknowledgment}
	We would like to thank N.~Evans, A.~O'Bannon, R.~Rodgers and H.~Furukawa for valuable comments and discussions. Specially thank R.~Rodgers for reading carefully the draft and comments. We are grateful to M.~J\"arvinen for the fruitful discussion and comments. This work is based upon research funded by Iran National Science Foundation (INSF) under project No. 99024938.  
	
	\begin{appendices}

	 \setcounter{equation}{0}  
	 \counterwithin*{equation}{section}
	 \renewcommand\theequation{\thesection\arabic{equation}}
	\section{Fluctuation equation with axial $b$ field} \label{B}
	 In this appendix, we give the fluctuation equation at finite momentum in the presence of the pseudoscalar field $ \psi(z)=bz $. Firstly, we give the embedding equation as	 
	 \begin{eqnarray}\label{embeddingb}
	 	0 &=& b^2u^3 \theta'(u)^2 \sin ^2\big(\theta(u)\big)~\cos \big(\theta(u)\big)\nn \\
	 	&&
	 	\qquad \bigg(2+2u^4+(2-q)u^q+(2+q)u^{4+q}\bigg) \nn \\
	 	&&
	 	+(1+u^q)\big(1+S(u)\big)\sin \big(\theta(u)\big) \nn \\
	 	&&
	 	\qquad \bigg(-1+u^2(-1+u^4)\theta'(u)^2\bigg) \nn \\
	 	&&
	 	+\frac{1}{2}~u~\cos \big(\theta(u)\big) \nn \\
	 	&&
	 	\qquad \Bigg((6+2u^4+2(3-q)u^q+2(1+q)u^{4+q})\theta'(u) \nn \\
	 	&&
	 	\qquad+u^2(-1+u^4)\bigg(-8+4u^4-(8+2q)u^q\nn \\
	 	&&
	 	\qquad+(4-2q)u^{4+q}+b^2u^2~L(u)~\big(1-\cos \big(2\theta(u)\big)\big)\bigg)\theta'(u)^3 \nn \\
	 	&&
	 	\qquad-u(-1+u^4)(1+u^q)~S(u)~\theta''(u)\Bigg). \nn \\
	 \end{eqnarray}
	 Next we apply the fluctuation as $ \Theta \equiv \Theta(t,z,u)=\theta (u)+\vartheta(t,z,u) $ where $\theta(u)$ is the solution of the embedding equation of motion \eqref{embeddingb}. Then the DBI action reads as	
	\begin{eqnarray}\label{actionfluc2}
		S_{D7} &=& \int dt~du  ~h(u) ~\frac{\cos ^3 \Theta}{u^5} \nn \\
		&& \sqrt{P  ~(\partial _u \Theta)^2+Q~(\partial _z \Theta)^2+W~\big(-1+u^4+u^2~(\partial _t \Theta)^2\big)},\nn \\
	\end{eqnarray}
	
	where
	
	\begin{eqnarray}\label{fluc12}
		P &=& -\frac{u^2}{2}(-1+u^4)~S(u),\nn \\
		Q &=&
		u^2,\nn \\
		W &=&
		-\big(1+b^2u^2 \sin ^2 \Theta \big).
	\end{eqnarray}
	
	 Assuming plane wave form $ \vartheta(t,z,u)=\text{e}^{-i(\omega t+kz)}\delta \theta(u) $ for the fluctuation, and keeping the equation of motion linearized, we find the fluctuation equation as  
	\begin{eqnarray}\label{fluc2}
		0 &=& \delta \theta ''(u)+F~\delta \theta'(u)+G~\delta \theta (u),
	\end{eqnarray}
	where
	\begin{eqnarray}\label{fluc3}
		G &=& \frac{\big(1-u^2(-1+u^4)~\theta'(u)^2\big)~\sec ^2(\theta(u))}{8u^2(-1+u^4)~S(u)^2}~H\nn \\
		F &=&
		\dfrac{\sec (\theta(u))}{2u(-1+u^4)(1+u^q)~S(u)}~K\nn \\
	\end{eqnarray}

	and
	
	\begin{eqnarray}\label{fluc4}
		S(u) &=& 2+b^2u^2\bigg(1-\cos \big(2\theta(u)\big)\bigg)\nn \\
	\end{eqnarray}
	
	also
	
	\begin{eqnarray}\label{fluc5}
		H &=&
		4(-1+u^4)\big(-24+k^2(4+b^2u^2)u^2\nn \\
		&& -u^2(22b^2-10b^4u^2)\big)+2u^2(8+4b^2u^2+b^4u^4)\omega^2\nn \\
		&&
		+u^2 \cos \big(2\theta(u)\big)\nn \\
		&&
		\qquad \big(16(7b^2+k^2+3b^4u^2)(-1+u^4)\nn \\
		&&
		\qquad \qquad+(16-b^4u^4)\omega^2\big)\nn \\
		&&
		+2b^2u^2 \cos \big(4\theta(u)\big)\nn \\
		&&
		\qquad \big(-2(-1+u^4)(-2+u^2(2b^2+k^2))\nn \\
		&&
		\qquad \qquad-u^2(4+b^2u^2)\omega^2\big)\nn \\
		&&
		+b^4u^6\omega^2 \cos \big(6\theta(u)\big)\nn \\
		&&
		+8b^2u^3(-1+u^4)^2 \theta'(u)\nn \\
		&&
		\qquad \bigg(2\sin\big(2\theta(u)\big)+\sin\big(4\theta(u)\big)\bigg)\nn \\
		\end{eqnarray}
		\begin{eqnarray}\label{fluc6}
		K &=&
		4\cos \big(\theta(u)\big)~\big(-L(u)+2 u^4(1+u^q)\big)\nn \\
		&&
		+b^2u^2~\cos \big(3\theta(u)\big)\nn \\
		&&
		\qquad \big(2+2u^4+u^q(2-q+(2+q)u^4)\big)\nn \\
		&&
		+\theta'(u)(1+u^q)\sin \big(\theta(u)\big)(-1+S(u))\nn \\
		&&
		+6u^2\theta'(u)^2(-1+u^4)\cos \big(\theta(u)\big)\nn \\
		&&
		\qquad \Bigg(-8+4u^2-2u^q(4-q+u^4(q-2))\nn \\
		&&
		\qquad +b^2u^2~L(u)\bigg(1-\cos \big(2\theta(u)\big)\bigg)\Bigg)\nn \\
	\end{eqnarray}
	
	with
	
	\begin{eqnarray}\label{fluc6}
		L(u) &=& -3+u^4+u^q\big(-3+u^4+q(-1+u^4)\big).\nn \\
	\end{eqnarray}

\end{appendices}


\end{document}